\documentclass[acmsmall]{acmart}
\usepackage{algorithm}
\usepackage[noend]{algpseudocode}
\usepackage{enumitem}
\usepackage{fbox}
\usepackage{graphicx}
\usepackage{hyphenat}
\usepackage{listings}
\usepackage{multirow}
\usepackage{tcolorbox}
\usepackage[normalem]{ulem}
\usepackage{wrapfig}
\usepackage[table,xcdraw]{xcolor}

\usepackage{soul,xcolor}
\setstcolor{red}

\useunder{\uline}{\ul}{}
\algrenewcommand\algorithmicensure{\textbf{Return:}}
\definecolor{ballblue}{rgb}{0.13, 0.67, 0.8}

\newcommand{\rectangle}{{%
\ooalign{$\sqsubset\mkern3mu$\cr$\mkern4mu\sqsupset$\cr}%
}}

\definecolor{codegreen}{rgb}{0,0.6,0}
\definecolor{codegray}{rgb}{0.5,0.5,0.5}
\definecolor{codepurple}{rgb}{0.58,0,0.82}
\definecolor{backcolour}{rgb}{0.95,0.95,0.92}

\lstdefinelanguage{json}{
  morestring=[b]",
  morecomment=[l]{//},
  moredelim=[l][\color{black}]{:},
  moredelim=[s][\color{red}]{[}{]},
  stringstyle=\color{teal},
  keywordstyle=\color{black},
  commentstyle=\color{gray},
  showstringspaces=false,
}
\AtBeginDocument{%
  }

\setcopyright{cc}
\setcctype{by}
\acmDOI{10.1145/3808137}
\acmYear{2026}
\acmJournal{PACMSE}
\acmVolume{3}
\acmNumber{FSE}
\acmArticle{FSE130}
\acmMonth{7}
\acmSubmissionID{fse26mainb-p431-p}
\received{2026-02-16}
\received[accepted]{2026-03-24}

\begin{document}

\title{TORAI: Multi-source Root Cause Analysis for Blind Spots in Microservice Service Call Graph}

\author{Luan Pham}
\orcid{0000-0001-7243-3225}
\affiliation{%
  \institution{RMIT University}
  \city{Melbourne}
  \country{Australia}
}
\email{luan.pham@rmit.edu.au}

\author{Huong Ha}
\orcid{0000-0003-2463-7770}
\affiliation{%
  \institution{RMIT University}
  \city{Melbourne}
  \country{Australia}
}
\email{huong.ha@rmit.edu.au}

\author{Xiuzhen Zhang}
\orcid{0000-0001-5558-3790}
\affiliation{%
  \institution{RMIT University}
  \city{Melbourne}
  \country{Australia}
}
\email{xiuzhen.zhang@rmit.edu.au}

\author{Hongyu Zhang}
\orcid{0000-0002-3063-9425}
\affiliation{%
  \institution{Chongqing University}
  \city{Chongqing}
  \country{China}
}
\email{hyzhang@cqu.edu.cn}

\begin{abstract}
Existing multi-source root cause analysis (RCA) methods for microservice systems assume all services have traces to construct a service call graph. However, this assumption is not practical as microservice systems evolve rapidly and may contain blackbox services without traces, such as compiled software or unsupported services. We refer to these services as \textit{blind spots}. In the presence of blind spots, the performance of existing multi-source RCA methods may be affected, as they only diagnose \textit{visible} services on the call graph. To overcome this limitation, we propose TORAI, a novel unsupervised approach that effectively pinpoints fine-grained root causes without relying on the service call graph. Instead, TORAI first measures anomaly severity using available multi-source telemetry data. It then performs clustering to group services based on their severity symptoms and conducts causal analysis to rank services within each severity cluster. Finally, TORAI aggregates the cluster rankings and uses hypothesis testing to identify fine-grained root causes. TORAI provides an unsupervised approach that leverages available multi-source telemetry data for RCA without requiring a constructed service call graph or further intrusive actions, thus addressing the limitations of existing methods. Our experiments on three benchmark systems demonstrate that TORAI outperforms state-of-the-art baselines remarkably in the presence of blind spots. Performance on real-world failures further shows that TORAI can accurately pinpoint the root causes in top-3 recommendations.

\end{abstract}




\setcopyright{rightsretained}
\copyrightyear{2026}

\maketitle

\section{Introduction} \label{sec:introduction}

Microservices have emerged as a popular form of software systems because of their advantages such as loose coupling, resource flexibility, and simplified deployment processes. However, the inherent complexity of microservice systems often leads to failures, affecting user experience, and causing substantial economic losses. For example, a one-hour downtime on Amazon may cost up to 100 million dollars~\cite{pham2024baro, pham2026eventadl, pham2026graph}. Therefore, effectively and efficiently diagnosing the root causes of failures in microservice systems is crucial to quickly mitigate the failure and minimize its impact.

To effectively conduct root cause analysis (RCA), system operators typically gather three primary sources of telemetry data: metrics, logs, and traces~\cite{lee2023eadro, yu2023nezha, zhang2023diagfusion}. Metrics, typically presented as time series, provide insights into service status, including system-level and application-level metrics like CPU usage and average response time. Logs, comprising semi-structured text, record events that occur at runtime, including request handling events, state changes. Traces, structured in a tree format, record the sequence of user request invocations. Unlike metrics and logs, traces usually require heavy instrumentation with distributed tracing, often involving modifications to microservices' source code~\cite{janes2023open, shen2023deepflow, ashok2024traceweaver, yu2023nezha}. Figure~\ref{fig:main}A illustrates the multi-source telemetry data.

\begin{wrapfigure}{r}{0.5\textwidth}
\vspace{-5pt}
\includegraphics[width=\linewidth]{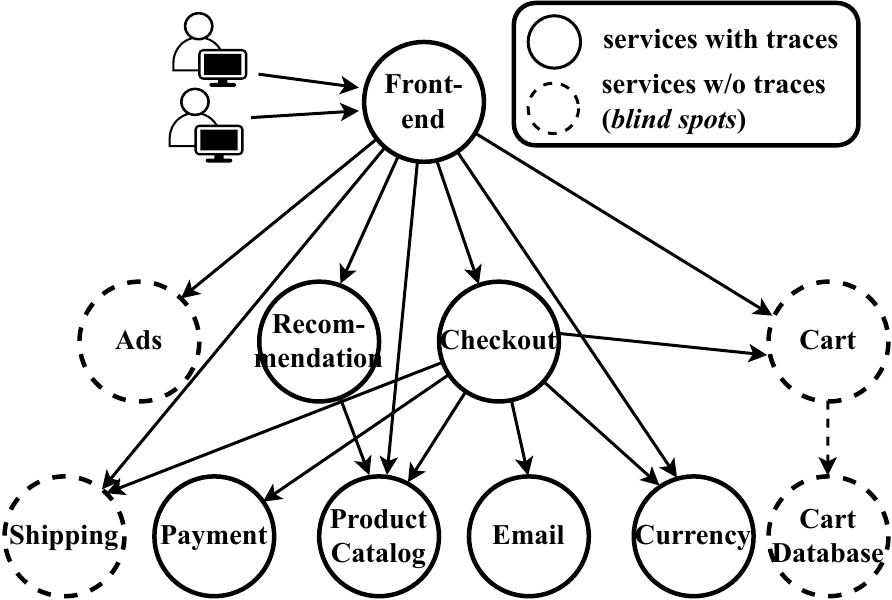}
\vspace{-15pt}
\caption{Service Call Graph of the Online Boutique microservice system containing \textit{blind spots} (i.e., services without distributed tracing instrumentation).}\label{fig:missing-traces-1}
\vspace{-5pt}
\end{wrapfigure} 
We have identified three key limitations of existing multi-source RCA methods~\cite{lee2023eadro, zhang2023diagfusion, rouf2024instantops, xie2024tvdiag, yu2023nezha}: they are either \textit{(1) approaches that assume the system has full trace coverage, allowing them to construct a full service call graph from traces, or (2) supervised approaches requiring a large volume of labeled data, or (3) intrusive approaches requiring heavy instrumentation of the microservices' source code}.
First, most multi-source RCA methods~\cite{lee2023eadro, zhang2023diagfusion, rouf2024instantops, xie2024tvdiag, yu2023nezha}, except those omitting trace information~\cite{zheng2024mulan, zhang2021cloudrca}, assume the system has full trace coverage, i.e., all services in the system are instrumented with distributed tracing, because they rely on traces to construct the service call graph required in their RCA methods. This requirement limits their applicability, especially in large and evolving microservice systems where new services or versions are introduced all the time, the engineers may not have enough effort to implement the distributed tracing for newly introduced services~\cite{shen2023deepflow, ashok2024traceweaver}, resulting in many \textit{blind spots} (i.e., services are not in the service call graph), that directly impact RCA performance, as shown in Figure~\ref{fig:missing-traces-1}.  Second, many multi-source RCA approaches ~\cite{lee2023eadro, zhang2023diagfusion, liu2022microcbr, xie2024tvdiag} are supervised approaches, requiring a large volume of training data to train their models. It is generally impractical to expect such a large volume of training data, as microservice systems are typically large and rapidly developed with many updates~\cite{shen2023deepflow, uber}. Third, a recent multi-source RCA method~\cite{yu2023nezha} requires a tight integration between traces and logs (i.e., each log line must contain the \texttt{trace\_id}), which demands extensive
\begin{wrapfigure}{r}{0.59\textwidth}
\vspace{-12pt}
\begin{tcolorbox}[left=2pt,right=2pt,top=0pt,bottom=0pt]
\textbf{Q: How do we address the mentioned limitations?}

\textit{We aim to propose a novel multi-source RCA method that satisfies three criteria: (1) avoiding using the call graph constructed from traces; (2) relying on unsupervised techniques; and (3) leveraging available telemetry data without requiring further updates 
to the source code.}
\end{tcolorbox}
\vspace{-10pt}
\end{wrapfigure} 

\noindent  effort to implement the necessary changes in the microservices' source code and may become infeasible when there are closed-source components. Consequently, these three limitations make existing multi-source RCA methods inflexible and challenging to implement in real-world scenarios.

In this study, we introduce TORAI, a novel multi-source RCA method for identifying fine-grained root causes of microservice failures. TORAI employs unsupervised techniques intuitively. First, it measures the anomaly severity from multi-source telemetry data. Second, it clusters services with similar severity symptoms. Next, it applies causal analysis to identify coarse-grained root causes (i.e., root cause services). Finally, it uses hypothesis testing to derive fine-grained root causes indicators (e.g., metrics or logs indicating the underlying root cause). TORAI offers several advantages over existing multi-source RCA methods. First, it does not rely on the service call graph (see Figure~\ref{fig:missing-traces-1}), eliminating the need for full trace coverage. Second, it uses unsupervised techniques such as clustering, and causal analysis, removing the need for labeled data and enabling direct application to microservice systems without requiring training. Third, it effectively leverages multi-source telemetry data, allowing for the root cause diagnosis in services not present in the call graph, without the need for additional intrusive instrumentation to achieve full trace coverage.

To evaluate TORAI, we conduct extensive experiments on 270 failure cases collected from three benchmark microservice systems and 10 real-world incidents, comparing TORAI against nine state-of-the-art RCA methods. The experimental results demonstrate that TORAI outperforms all baselines in localizing both coarse-grained and fine-grained root causes in terms of effectiveness and efficiency. We further evaluate TORAI on real-world failures from a major internet provider. The results show that TORAI outperforms state-of-the-art baselines and achieves 100\% accuracy in ranking the root cause within the top three recommendations. 

\noindent In summary, our main contributions are as follows: 
\begin{itemize}[leftmargin=*]
    \item We identify three key limitations of existing multi-source RCA methods, including the assumption of full trace coverage (i.e., assuming the constructed service call graph is complete and contains no \textit{blind spots}), the reliance on labeled data, and the requirement for intrusive instrumentation. These limitations motivate our work.
    \item We introduce TORAI, a novel RCA method that addresses the mentioned limitations effectively. TORAI pinpoints fine-grained root causes by diagnosing available multi-source telemetry without requiring a constructed call graph or labeled data, making it adaptable to a wide range of microservice systems. 

    \item We conduct extensive experiments to evaluate TORAI and the experimental results demonstrate that TORAI outperforms existing methods in both effectiveness and efficiency. Additionally, we evaluate TORAI with real-world failures and the results confirm its potential.
\end{itemize}

\section{Background}

\subsection{Key Terminologies}

\textit{Failures} denote the incapacity of a service to perform its functions~\cite{Soldani2022rcasurvey, pham2026eventadl}. \textit{Faults} correspond to the root causes of such failures (e.g., CPU overload, memory leaks, or network disconnections)~\cite{Soldani2022rcasurvey, avizienis2004basic}. \textit{Anomalies} are defined as observable symptoms of failures~\cite{pham2024baro, pham2026eventadl}. \textit{Root cause analysis (RCA)} is the process of identifying why a failure has occurred~\cite{lee2023eadro}, i.e., pinpointing the failure's root causes.
RCA entails a comprehensive examination of multi-source telemetry data (i.e., metrics, logs, and traces) to derive \textit{coarse-grained root causes} (i.e., root cause services), and \textit{fine-grained root causes} (i.e., root cause indicators).
The system operators can use these suggested root cause services and indicators (e.g., specific metrics, logs, or traces) to identify the underlying root cause of the failures. The use of terminologies aligns with existing RCA works~\cite{Azam2022rcd, Li2022Circa, Xin2023CausalRCA, pham2024baro}.

\subsection{Problem Formulation} \label{sec:problem-formulation}

Consider a microservice system $\mathcal{S}$ comprising $N$ services $\{s^i\}_{i=1}^N$. Multi-source telemetry data, including metrics, logs, and traces, are collected for each service, if available. At each time step $t$, we obtain multi-source telemetry data $\mathcal{D}_t^i=\{\mathcal{M}_t^{(i,m)}, \mathcal{L}_t^{(i,l)}, \mathcal{TC}_t^{(i,s)}\}$ for each service $s^i$, where $\mathcal{M}^{(i,m)}$ is a set of $m$ metrics, $\mathcal{L}_t^{(i,l)}$ represents a set of $l$ logs, and $\mathcal{TC}_t^{(i,s)}$ denotes a set of $s$ traces. Given the anomaly detection time $\hat{t}_A$ (i.e., the start timestamp of the abnormal window detected by the anomaly detection module), let us denote $\mathcal{D}_{t_0 \le t < \hat{t}_A}$ as the multi-source telemetry data collected during normal periods and $\mathcal{D}_{\hat{t}_A \le t < \hat{t}_A + T}$ as the data collected during the failure occurrence period, our objective is to develop a method that identifies the root causes of the failure using these datasets. The expected RCA output is a ranked list of root cause services and their corresponding root cause indicators (e.g., specific metrics, logs, or traces that indicate the root causes).

\subsection{Related Work and Motivation} \label{sec:related-world-and-motivation}

\subsubsection{The Blind Spots of Existing Multi-Source RCA} \label{sec:motivation-1-full-trace}
Existing multi-source RCA methods~\cite{yu2023nezha, lee2023eadro, zhang2023diagfusion, hou2021pdiagnose, rouf2024instantops, xie2024tvdiag} typically construct a service call graph from traces, assuming the system has full trace coverage, i.e., 100\% of services are instrumented with distributed tracing \cite{shen2023deepflow, ashok2024traceweaver}. They then integrate metrics and logs into the service call graph to perform RCA. This approach is only effective when the system has no \textit{blind spots}, i.e., services without traces or closed-source components \cite{shen2023deepflow, lee2023eadro, yu2023nezha, ashok2024traceweaver}. In this paper, we refer to services or components not detected by the traces but that should be as blind spots. In the presence of blind spots, existing multi-source RCA methods \cite{yu2023nezha, lee2023eadro, zhang2023diagfusion, gu2023trinityrcl, xie2024tvdiag} may not perform RCA effectively, as the constructed service call graph is incomplete. Consequently, key metrics and logs may be overlooked, and their behavior may not be analyzed when diagnosing the root cause of failure in microservice systems. Recent studies~\cite{shen2023deepflow, ashok2024traceweaver} and industry reports~\cite{odigos2025pitfalls} show that requiring the microservice systems to be fully instrumented is ineffective and inefficient. In reality, microservice systems are often developed in many different languages, and the call graph can become exceedingly intricate, with some systems comprising up to 1,500 services \cite{luo2022depth}. Shen et al.~\cite{shen2023deepflow} show that engineers spend hours to instrument mere tens of lines of code for a single component. Hence, they may not have sufficient time to instrument every component before deployment, leading to numerous blind spots in the constructed call graph. TraceWeaver~\cite{ashok2024traceweaver} explicitly acknowledges that eBPF is ``insufficient to solve the tracing challenge'' as ``there is no guarantee that the application will propagate these headers to related outgoing backend requests.'' Industry reports~\cite{odigos2025pitfalls} confirm that blind spots remain prevalent in production systems despite advances in eBPF-based tracing. In this paper, we aim to bypass the assumption of full traces coverage and develop a  multi-source RCA method to localize root causes effectively, even when some or all traces are missing. It is important to note that metrics and logs are relatively easy to collect, as they do not require source code modification, unlike traces. Metrics such as response time and error rates can be collected without traces by monitoring agents or service mesh without requiring any source code modification~\cite{pham2024baro, shen2023deepflow}. This is standard practice in industry~\cite{googlesre} and is how existing metric-based RCA works collect data~\cite{pham2024baro, Azam2022rcd, Li2022Circa}.

\vspace{-5pt}
\subsubsection{Supervised RCA Approaches In A Dynamic World} \label{sec:motivation-unsupervise}
Supervised multi-source RCA methods~\cite{lee2023eadro, zhang2023diagfusion, zhang2021cloudrca, xie2024tvdiag} rely on large volumes of manually labeled training data to train their models, which often incorporate graph neural networks or convolutional layers. However, requiring such large amounts of labeled data is often impractical or challenging in real-world scenarios. First, microservice systems are highly dynamic, with old services being updated and new services being released frequently. Second, historical faults are often resolved, while new faults are introduced over time~\cite{shen2023deepflow}. Additionally, future edge-case faults, which could cause significant losses, are unlikely to be present in historical failure datasets~\cite{zhang2023benefit}. Third, microservice systems are typically large in scale~\cite{uber}, making it costly and impractical to manually label data to cover all services and fault types. Therefore, our aim is to develop an RCA method that leverages unsupervised techniques to reduce the dependency on labeled data.

\vspace{-5pt}
\subsubsection{Tight Integration between Multi-source Data} \label{sec:motivation-2-independent}
Recent multi-source RCA studies~\cite{yu2023nezha, zhang2023diagfusion, lee2023eadro, zheng2024mulan, zhang2021cloudrca} require multi-source telemetry data to be tightly integrated. For example, Nezha~\cite{yu2023nezha} requires every log line in the microservices to be modified to contain a \texttt{trace\_id}. To achieve this, engineers must manually modify every log or print statement in the microservices' source code, which is time-consuming and costly. This task becomes impossible when dealing with black box components, third-party services, or new frameworks that do not yet support distributed tracing~\cite{giamattei2023monitoring, shen2023deepflow}. In addition, Eadro \cite{lee2023eadro} and DiagFusion \cite{zhang2023diagfusion} extract features from metrics and logs, integrate these features into a graph constructed using traces. These methods struggle in the presence of blind spots, as many metrics and logs may not correspond to any location on the graph, leading to the failure to use these valuable data for failure diagnosis. In summary, most existing multi-source RCA approaches may be impractical, as their assumptions often might not be fully met in real-world scenarios. Some other multi-source RCA methods do not require this tight integration, but interestingly, they completely omit trace information \cite{zhang2021cloudrca, zheng2024mulan, wang2020root} or log information \cite{zhu2024hemirca}. In other words, they fail to fully take advantage of all available telemetry data. 

\vspace{-5pt}
\subsubsection{Summary} These motivations drive the design of TORAI, a novel multi-source RCA method that overcomes the existing limitations. Firstly, our method leverages multi-source telemetry data without requiring full trace coverage \cite{yu2023nezha, lee2023eadro, zhang2023diagfusion}, allowing it to diagnose the fine-grained root causes of failures with high performance even in the absence of some or all traces. We do not attempt to reconstruct the call graph, instead, we design TORAI to perform RCA effectively without requiring a complete call graph. This design is inspired by recent theoretical work~\cite{neurips2025rca}, which proves that identifying root causes based on anomaly scores is causally justified. Recent studies~\cite{pham2024root} also show that call graph reconstruction can be ineffective for RCA due to limitations of causal discovery algorithms. Secondly, our method can be applied without modifying the source code of microservices for tight integration between traces and logs~\cite{yu2023nezha}. Thirdly, our method does not rely on labeled training data \cite{lee2023eadro, zhang2023diagfusion, zhang2021cloudrca, xie2024tvdiag}, making it adaptable to a wide range of systems. The details of our proposed RCA method, TORAI, are presented in the next section.

\section{TORAI: Unsupervised Fine-grained RCA using Multi-Source Telemetry Data}

When an anomaly is detected, TORAI is triggered to perform RCA as follows. First, TORAI collects the multi-source telemetry data and transforms them into time series (Sec. \ref{sec:method-collect}). Second, it measures the anomaly severity of each data source for all services using \textit{SeverityScorer} (Sec.~\ref{sec:severity-scorer}). Third, it uses \textit{SymptomCluster} to group services exhibiting similar severity symptoms (Sec.~\ref{sec:symptom-cluster}). Fourth, it conducts causal analysis to rank the root causes within each severity group using \textit{CausalRanker} (Sec. \ref{sec:causal-ranker}). Then, it performs \textit{RankAggregation} to aggregate the results from the SymptomCluster and CausalRanker steps, yielding a root cause service ranked list (Sec.~\ref{sec:rank-aggregation}). Finally, TORAI uses \textit{FineGrainer} to perform hypothesis testing and derive fine-grained root cause indicators for the corresponding root cause services (Sec. \ref{sec:fine-grainer}). The overview of TORAI is shown in Fig.~\ref{fig:main}.

\subsection{Transform Multi-source Telemetry Data} \label{sec:method-collect}

TORAI transforms multi-source telemetry data into time series for RCA. It collects data during normal periods to learn expected behaviors and during abnormal periods to analyze and identify root causes. The processing steps for each data source are as follows:

\subsubsection{Metrics}
TORAI collects four metrics types: \textit{Traffic} (e.g., requests per minute), \textit{Saturation} (e.g., CPU/memory utilization), \textit{Latency} (e.g., average response time), and \textit{Errors} (e.g., the rate of failed requests)~\cite{pham2024baro}, known as the four golden signals in site reliability engineering~\cite{googlesre}, during both normal and abnormal periods. For each service $s^i$, TORAI compiles a set of time series $\mathcal{X}_\mathcal{M}^i$ to represent the metrics data.

\subsubsection{Logs}
We focus on log occurrences rather than log semantics, as empirical studies have shown that the quality of log semantics cannot be guaranteed~\cite{he2021survey, lee2023eadro} and semantic extraction requires significant computational resources. First, TORAI parses logs into log templates using Drain \cite{he2017drain} by removing variables from the log messages. Next, TORAI bins the occurrence of log templates based on their timestamps to obtain time series. This process generates a set of time series $\mathcal{X}_\mathcal{L}^i$, representing the log occurrences of each service $s^i$. 

\subsubsection{Traces.}
We extract available information from traces, but do not construct the call graph as it may be incomplete (see Sec. ~\ref{sec:related-world-and-motivation}). TORAI transforms latency and status code in traces into time series. Each time series presents the frequency of latency and status code for each trace operation.

We refer to time series data derived from multi-source telemetry data as \textbf{multi-source time series data}. Each service $s^i$ has a set of multi-source time series $\mathcal{X}^i = \{ \mathcal{X}^i_\mathcal{M}, \mathcal{X}^i_\mathcal{L}, \mathcal{X}^i_\mathcal{TC} \}_{t_0 \le t < \hat{t}_A +T}$ where $\mathcal{X}^i_\mathcal{M}$, $\mathcal{X}^i_\mathcal{L}$, and $\mathcal{X}^i_\mathcal{TC}$ denote the set of time series for metrics, logs, and traces. The data is split at the anomaly detection time $\hat{t}_A$ into normal ($\mathcal{X}^i_{t_0\le t<\hat{t}_A}$) and abnormal ($\mathcal{X}^i_{\hat{t}_A\le t< \hat{t}_A+T}$) sets.

\begin{figure}[t]
\centering
\includegraphics[width=\textwidth]{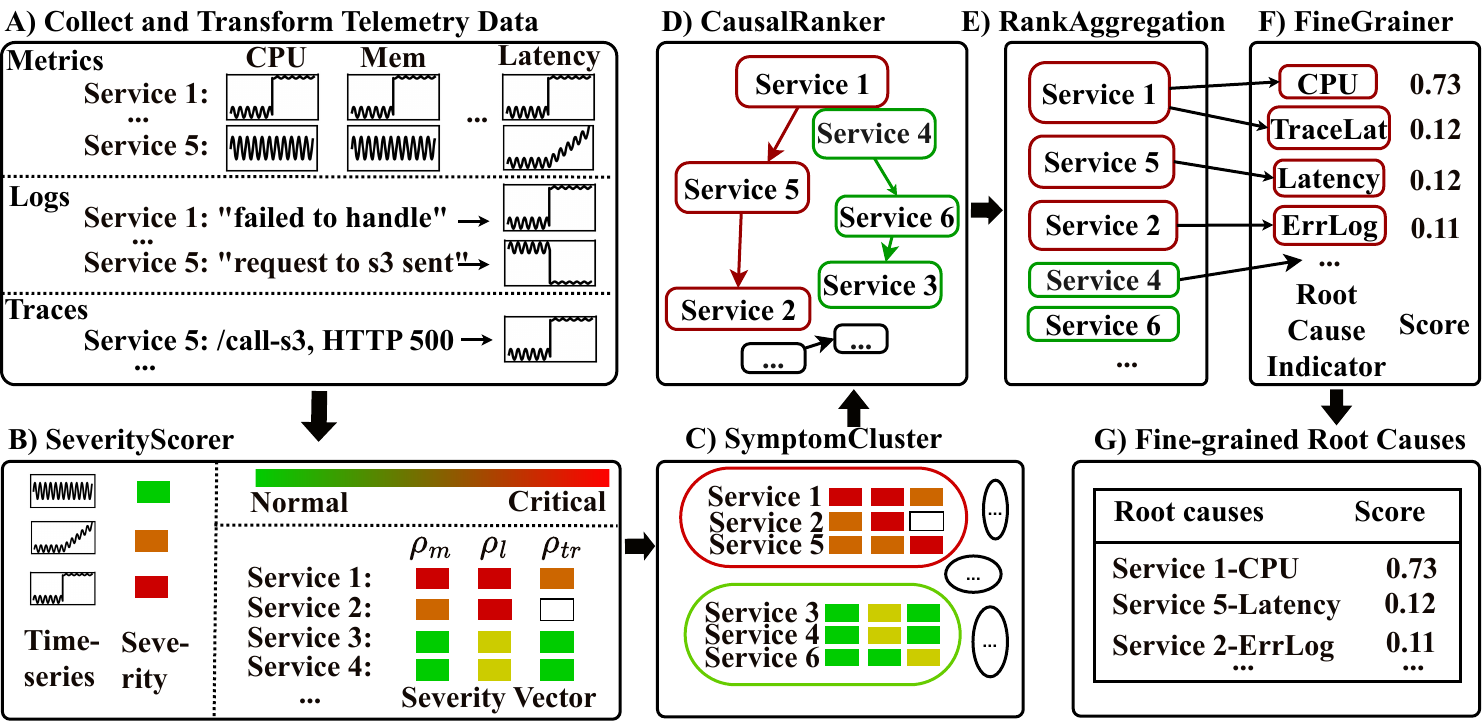}
\vspace{-10pt}
\caption{Overview of TORAI. (A) TORAI transforms telemetry data into time series. (B) It computes anomaly severity scores, producing vectors [$\rho_m^i, \rho_l^i, \rho_{tc}^i$] for each service $s^i$ (with missing data sources denoted by $\protect\rectangle$; e.g., Service 2 is a \textbf{blind spot}). (C) TORAI clusters services based on their severity symptoms. (D) Within each cluster, it applies causal inference-based RCA to rank potential root causes. (E) TORAI aggregates rankings from steps C and D to generate a coarse-grained root cause list. (F) It then conducts fine-grained RCA using hypothesis testing to produce (G) a fine-grained ranked list of root causes.} \label{fig:main}
\vspace{-17pt}
\end{figure}

\subsection{SeverityScorer} \label{sec:severity-scorer}

SeverityScorer measures anomaly severity scores for each time series and generates a severity vector for each service. These severity vectors effectively assist SymptomCluster in grouping services based on their severity symptoms, thereby quickly separating abnormal services from normal ones (see Sec.~\ref{sec:symptom-cluster}). This also reduces overhead for CausalRanker, which further analyses the multi-source time series to derive the root causes (see Sec.~\ref{sec:causal-ranker}). It is important to note that SeverityScorer does not aim to identify the root causes directly but instead focuses on quickly assessing the severity of each service based on its multi-source telemetry data after a failure occurs, thereby helping subsequent components of TORAI concentrate their focus on analysing the abnormal services.

In particular, for each time series $x^{(i, j)}$, SeverityScorer learns the mean $\mu^{(i, j)}$ and standard deviation $\smash{\sigma^{(i, j)}}$ during the normal period. Then during the abnormal period, for each data point $\smash{x_t^{(i, j)}}$ of the time series $x^{(i, j)}$, SeverityScorer measures how far it diverges from the expected value. This deviation is denoted as $a^{(i, j)}_t$ and is computed as $\smash{a^{(i, j)}_t = \big | x^{(i, j)}_t - \mu^{(i, j)} \big | / \sigma^{(i, j)}}$.  SeverityScorer then aggregates $a^{(i, j)}_{t}$ for all the available data during the abnormal period, yielding the anomaly severity score $\smash{\rho^{(i, j)} = \max_{\hat{t}_A \leq t < \hat{t}_A + T} a^{(i, j)}_t}$.

Using the anomaly severity scores ${\rho^{(i, j)}}$ across all time series and data sources, SeverityScorer assigns each service $s^i$ a severity vector $[\rho_m^i$,~$\rho_l^i$,~$\rho_{tc}^i]$, which represents the severity status of each service as observed through its multi-source telemetry data. If a data source is missing, SeverityScorer imputes the corresponding $\rho$ value as 0, indicating no anomalies were detected from that source. This flexible design enables TORAI to use available data sources, such as metrics and logs, without requiring the presence of traces. Specifically, $\rho_m^i$ and $\rho_{tc}^i$ are obtained by integrating the severity scores of the time series derived from metrics and traces associated with service $s^i$, i.e.,  $\rho_m^i = \sum_{x^{(i,j)}\in \mathcal{X}_\mathcal{M}^i} \rho_m^{(i,j)}$ and $\rho_{tc}^i = \sum_{x^{(i,j)}\in \mathcal{X}_\mathcal{TC}^i} \rho_{tc}^{(i,j)}$. For logs, $\rho_l^i$ is determined as the score of the most anomalous log templates for $s^i$, i.e., $\smash{\rho_{l}^i = \max_{x^{(i,j)}\in \mathcal{X}_\mathcal{L}^i} \rho_{l}^{(i,j)}}$, since the number of log templates between services varies significantly.

\subsection{SymptomCluster} \label{sec:symptom-cluster}

SymptomCluster performs clustering to group services with similar severity symptoms. The goal is to effectively separate abnormal services from normal ones, reducing the overhead for CausalRanker when conducting causal analysis, inspired by recent research~\cite{pham2024root} that highlights how including all services in the causal analysis can be inefficient and ineffective.

Specifically, SymptomCluster proposes using Gaussian Mixture Model (GMM) \cite{gmm} for this clustering task. GMM is a probabilistic model that assumes the data come from a mixture of several Gaussian distributions, each representing a cluster. GMM is known for its efficiency and has been successfully applied in various domains~\cite{patel2020clustering}. We first use the Bayesian Information Criterion (BIC)~\cite{Schwarz1978bic} to identify the optimal number of clusters (i.e., the number of Gaussian distributions). Specifically, we iterate through all possible values for the number of clusters. For each iteration, SymptomCluster runs GMM on the set of severity vectors $\{[\rho_m^i, \rho_l^i, \rho_{tc}^i]\}_{i=1}^N$ ($N$ is the number of services) and calculates the corresponding BIC score. The optimal number of clusters is chosen as the number that yields the lowest BIC score \cite{gmm, pham2024root}.

\setlength{\columnsep}{12pt}
\begin{wrapfigure}{r}{5cm}
\vspace{-20pt}
\includegraphics[width=\linewidth]{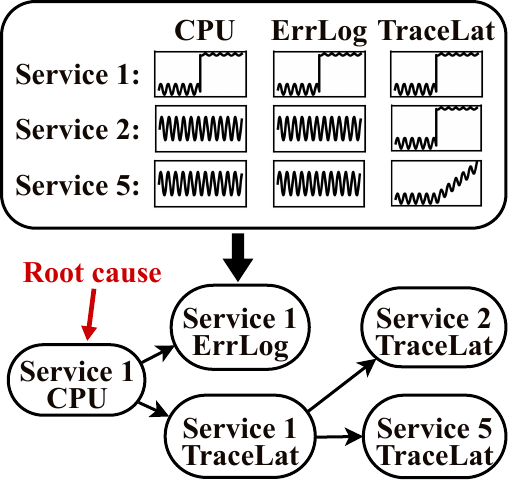}
\vspace{-15pt}
\caption{CausalRanker analyses the multi-source time series data of all services within each severity group to construct a causal graph and identify the root causes. (ErrLog: Error Logs, TraceLat: Latency extracted from Traces).}
\label{fig:causal-ranker}
\vspace{-30pt}
\end{wrapfigure}
Once the optimal number of clusters is determined, SymptomCluster performs clustering again using this value. For each cluster, SymptomCluster measures the cluster severity score as the mean severity score of all services within that cluster (i.e., the mean of the Gaussian distribution). Finally, SymptomCluster generates a ranked list of clusters (a cluster-level ranked list), where the highest-ranked clusters are the most likely to contain the root cause. This approach aligns with the intuition that operators should prioritize higher severity services for failure diagnosis, while setting lower priority to services with less severity and normal services. In addition, GMM allows soft clustering, meaning a service may be probabilistically assigned to multiple clusters. Consequently, a ``weak'' anomalous service can be included in multiple clusters for causal analysis, enabling the discovery of cross-cluster causal relationships.

\subsection{CausalRanker: Causal-based Root Cause Analysis} \label{sec:causal-ranker}
CausalRanker is designed to identify the root causes within each severity group obtained from the previous step. It operates on the belief that the root cause services may not exhibit the strongest anomaly severity but will have cause-effect relationships with affected services. These causal relationships are reflected in their time series data. CausalRanker performs causal inference-based RCA, analysing the multi-source time series of all services in each severity group by examining their anomalous patterns (e.g., periodicity, causality) to rank the root causes in each group.

Our CausalRanker ranks services within each cluster via divide-and-conquer causal inference using multi-source time series data. A causal graph is a directed acyclic graph (DAG) where nodes represent time series (e.g., CpuUsage, ErrLog) and directed edges represent causal relationships (e.g., CpuUsage $\rightarrow$ ErrLog indicates that CPU anomaly causes error log anomaly). This is a standard representation in causal discovery literature~\cite{Spirtes1995fci, Jaber2020PsiFCI}. A root cause is identified as a node that has no parents in the causal graph (i.e., an interventional target).

Specifically, CausalRanker randomly partitions the given set of time series into smaller groups called chunks. The time series are randomly partitioned into smaller chunks for efficient causal analysis, as analyzing all time series simultaneously is computationally expensive and has been shown to degrade performance~\cite{pham2024root}. It then applies the $\Psi$-PC \cite{Jaber2020PsiFCI}, an advanced causal discovery method in the presence of \textit{interventions} (i.e., failures), to construct causal graphs and identify root causes, referred to as interventional targets, within each chunk. Recursively, CausalRanker groups these root causes and repeats the process until only one chunk remains. Finally, the root causes are ranked based on their (conditional) independence test scores. Lower independence test scores indicate stronger causal influence, suggesting higher likelihood of being the root cause. This provides a ranked list of root cause services for each cluster.

Consider the example in Figure~\ref{fig:main}C and Figure~\ref{fig:causal-ranker}, ``Service~1", ``Service~2", and ``Service~5" are grouped in the same severity group. In this example, the root cause is ``Service~1~-~CPU", which causes an increase in error logs (Service 1-ErrLog) and latency (Service~1~-~TraceLat). Therefore, we design CausalRanker to construct the causal graph and identify ``Service~1" as the root cause service within this severity group. 

Our CausalRanker offers two key advancements over existing methods~\cite{Azam2022rcd, Xin2023CausalRCA}. First, prior studies~\cite{Azam2022rcd, Xin2023CausalRCA} use all available time series for causal analysis, which has recently been shown to degrade the performance of causal inference~\cite{pham2024root}. In contrast, CausalRanker leverages SymptomCluster, which groups services with similar severity levels, ensuring that only relevant time series are analyzed for causal relationships to derive the root cause. Second, previous studies rely solely on causal discovery algorithms to identify fine-grained root causes (e.g., RCD~\cite{Azam2022rcd} employs the \(\Psi\)-PC algorithm, while CausalRCA~\cite{Xin2023CausalRCA} uses DAG-GNN~\cite{Yu2019DagGNN}), an approach shown to be ineffective~\cite{pham2024root}. In TORAI, we introduce FineGrainer, which applies a robust hypothesis testing method to produce more accurate fine-grained root cause rankings within our framework (see Sec.~\ref{sec:fine-grainer}).

\subsection{RankAggregation} \label{sec:rank-aggregation}

In this phase, we combine the ranked lists from SymptomCluster and CausalRanker to produce an aggregated ranked list of coarse-grained root cause services.  We first prioritize the cluster-level ranking provided by SymptomCluster (Sec. \ref{sec:symptom-cluster}), meaning that services in clusters with higher severity are ranked higher. Second, within each cluster, we follow the ranking provided by CausalRanker (Sec. \ref{sec:causal-ranker}), meaning that if Service A is ranked higher than Service B within a cluster, then Service A will also be ranked higher than Service B in the final aggregated list. For instance, given a cluster-level ranking [Group A, Group B], where the internal ranking for Group A is [a,c] and for Group B is [e,p,t], our RankAggregation produces the corresponding aggregated ranking [a,c,e,p,t]. The highest-ranked items in the aggregated list have the highest probability of being the root cause service of the failure. The pseudo code for RankAggregation is presented in Algorithm \ref{alg:rank-aggregation}.

From our empirical observations, the top severity cluster may contain a single or multiple services. When the top severity cluster contains only one service, it is possible that the root cause service belongs to the second-highest severity cluster. Therefore, we apply CausalRanker to all clusters, not just the top severity cluster, and aggregate the results through this Rank Aggregation step.

In cases where the root cause does not exhibit strong anomalies (e.g., a code defect that only manifests in downstream services), TORAI ranks the affected services at the top. This allows quicker identification of actual root causes, instead of troubleshooting all possible services. For example, if a code defect causes correlated failures in services A and B, TORAI ranks A and B as top candidates, enabling operators to inspect these services and trace back to the true root cause promptly.

\begin{algorithm}[h]
\caption{Pseudo-code of RankAggregation} \label{alg:rank-aggregation}
\begin{algorithmic}[1]
\Require a ranked list $R_{cluster}$ from SymptomCluster (Sec. \ref{sec:symptom-cluster}), function CausalRanker (Sec. \ref{sec:causal-ranker}) 
\Ensure a ranked list of coarse-grained root cause services $R_s$

\Function{RankAggregation}{$R_{cluster}$, CausalRanker}
  \State $R_s \gets$ empty list
  \For{$cluster \in R_{cluster}$} \quad // loop from the highest to lowest ranked cluster
    
    \quad // get the corresponding ranked list from \textit{CausalRanker} of this \textit{cluster}
    \State $R_i \gets $ CausalRanker(\textit{cluster}) 
    \For{$s \in R_i$}
        \State $R_s \gets R_s$ appends $s$
    \EndFor 
  \EndFor 
  \State \Return $R_s$
\EndFunction
\end{algorithmic}
\end{algorithm}

\begin{figure}
\vspace{-10pt}
\centering
\begin{minipage}{0.45\linewidth}
\includegraphics[width=\linewidth]{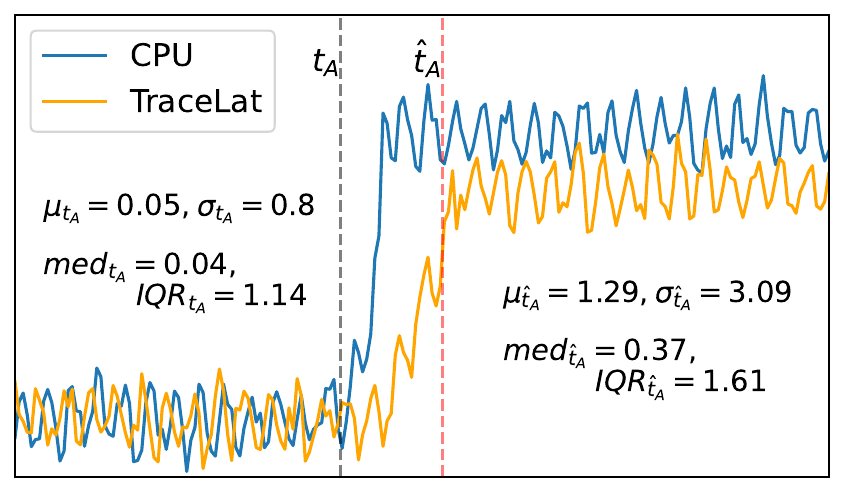}
\end{minipage}%
\hfill
\begin{minipage}{0.52\linewidth}
\vspace{-10pt}
\caption{\textbf{The Robustness of FineGrainer to Imprecise Anomaly Detection.} At time $t_A$, a failure occurs, causing a spike in CPU that eventually leads to increased latency (TraceLat). At $\hat{t}_A$, TraceLat surpasses the anomaly detection threshold, triggering an anomaly detection. This delayed detection introduces abnormal data (outliers) into the normal period of the CPU time series. The median and interquartile range (IQR) demonstrate greater robustness to these outliers compared to the $\mu$ and $\sigma$.}\label{fig:fine-grainer}
\end{minipage}
\vspace{-15pt}
\end{figure}

\vspace{-15pt}
\subsection{FineGrainer} \label{sec:fine-grainer}

After identifying the coarse-grained root cause services, we propose FineGrainer to determine fine-grained root cause indicators (e.g., specific metrics or logs indicating the root cause of failure) for each service, which better assists operators in diagnosing underlying issues~\cite{Li2022Circa, pham2024baro, liu2023pyrca}.

FineGrainer performs hypothesis testing on time series indicators to detect significant distribution changes after the anomaly detection time~\cite{Li2022Circa, pham2024baro}. Our key insight is that root cause indicators exhibit significant distributional changes after a failure occurs. 
We perform hypothesis testing where the null hypothesis $H_0$ states that the time series distribution remains unchanged after $\hat{t}_A$. 
 
Specifically, for the time series $x$ of each indicator (i.e., the time series data corresponding to each metric, log template, or trace), FineGrainer learns the median ($med$) and interquartile range ($IQR$) of the time series $x$ prior to the anomaly, spanning from $t_0$ to $\hat{t}_A$. It then measures how significantly each time series $x$ deviates from its expected central tendency during the failure period. This deviation is quantified as $a_t = \big | x_t - med \big | / IQR$. All values of $a_t$ across the time step $t$ during the failure period are consolidated to yield $\gamma = \max_{\hat{t}_A \leq t < \hat{t}_A + T} a_t$. Higher $\gamma$ values provide stronger evidence against $H_0$, indicating greater likelihood that the indicator is the root cause.

We prefer median and IQR over mean and standard deviation for analysing fine-grained root cause indicators. Prior work using mean and standard deviation~\cite{Li2022Circa} can degrade when anomaly detection times $\hat{t}_A$ are imprecise, as outliers from the actual failure period contaminate the normal period statistics. Median and IQR are more robust to such outliers (see Figure~\ref{fig:fine-grainer}).

\section{Experimental Results} \label{sec:experimental}

This section addresses the following research questions:
\begin{itemize}[leftmargin=*]
    \item RQ1: How effective is TORAI in coarse-grained RCA? 
    \item RQ2: How effective is TORAI in fine-grained RCA? 
    \item RQ3: How efficient is TORAI in performing RCA? 
    \item RQ4: How do TORAI's core components contribute to its overall performance?  
    \item RQ5: How does TORAI perform in real-world scenarios?
\end{itemize}

\subsection{Datasets} \label{sec:dataset}

% \begin{wraptable}{r}{7.85cm}
% \vspace{-14pt}
\begin{table}
\centering
\caption{Properties of collected datasets% from benchmark systems 
(\#service, \#metric, \#log, \#trace: number of services, metrics, log templates, and trace operations, per case. \#fault: number of fault types).} \label{tab:real-data}
\vspace{-5pt}
% \resizebox{\linewidth}{!}{%
\resizebox{9cm}{!}{%
\setlength\tabcolsep{1pt}
\begin{tabular}{l c c c c c c c c}
\hline
\textbf{Name} & \textbf{\#service} & \textbf{\#metric} & \textbf{\#log} & \textbf{\#trace} & \textbf{\#fault} & \textbf{\#cases} \\ \hline \hline
Online Boutique & 11 & 77 & 33$\pm$9 & 17  & 6 & 90  \\ \hline
Sock Shop  & 11 & 74 & 67$\pm$53 & - & 6 & 90   \\ \hline
Train Ticket & 64 & 376 & 163$\pm$44 & 148.3$\pm$26 & 6 & 90  \\ \hline
\end{tabular}%
}
\vspace{-12pt}
\end{table}

We deploy three widely used microservice benchmark systems, namely Online Boutique \cite{ob}, Sock Shop \cite{sockshop}, and Train Ticket \cite{tt}, on a Kubernetes cluster featuring five worker nodes, all configured with their default settings. Online Boutique is an e-commerce platform comprising 11 services that facilitate tasks such as browsing items, adding items to a user's cart, and making orders. Sock Shop, another online shopping system, consists of 11 services communicating via HTTP. Train Ticket is a ticket booking system that has 64 services, making it one of the largest benchmark microservice systems. Compared to Online Boutique and Sock Shop, Train Ticket has a complex design, various types of invocations and many log templates. The three benchmark microservice systems are widely used to evaluate RCA performance in existing literature~\cite{Azam2022rcd, wu2021microdiag, Xin2023CausalRCA, he2022graph, dan2021tracerca, yu2021microrank, zhou2018trainticket, Wang2021evalcausal}.

\begin{wrapfigure}{r}{8.5cm}
\vspace{-4pt}
\includegraphics[width=\linewidth]{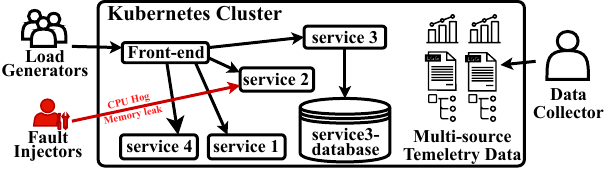}
\vspace{-16pt}
\caption{Illustration of our setup for the microservice systems and the multi-source telemetry data collection.} \label{fig:experiment-setup}
\vspace{-8pt}
\end{wrapfigure}
To simulate user interactions, we customize the provided load generators of these systems to emulate a load of 40-50 requests per second across all services. To gather metrics, we employ Prometheus \cite{prometheus}, along with cAdvisor \cite{cadvisor} and the Istio service mesh \cite{istio}, to monitor and collect application-level and resource-level metrics. For log collection, we deploy Datadog Vector \cite{vector} and Loki \cite{loki} to gather and aggregate logs from all service instances, storing them in Elasticsearch \cite{elasticsearch}. Traces are gathered using Jaeger \cite{jaeger}, with data sent to Elasticsearch for storage. Our telemetry data collection setup is depicted in Fig. \ref{fig:experiment-setup},
similar to existing works~\cite{Azam2022rcd, Xin2023CausalRCA, wu2021microdiag, wu2022automatic, yu2023nezha, lee2023eadro, Jinjin2018Microscope, he2022graph, dan2021tracerca, yu2021microrank, zhou2018trainticket, Wang2021evalcausal}.

We inject six common faults: CPU hog (CPU), memory leak (MEM), disk IO stress (DISK), socket stress (SOCKET), network delay (DELAY), and packet loss (LOSS) into key services of each benchmark system. In particular, we inject faults into five Sock Shop services (user, catalogue, orders, payment, and carts), five Online Boutique services (email, currency, recommendation, product, checkout), and five Train Ticket services (order, route, auth, train, travel). These services play an important role in their respective systems, as issues with their performance quickly impact the overall health of the system \cite{Azam2022rcd, Xin2023CausalRCA, lee2023eadro}. Firstly,  we let the systems run normally for ten minutes to gather normal metrics, logs, and traces. Then, we follow the existing practice \cite{Azam2022rcd, lee2023eadro, wu2021microdiag, Xin2023CausalRCA, yu2021microrank} to inject faults into the running services. For CPU, MEM, DISK, and SOCKET, we use stress-ng \cite{stressng} to stress the container resource. For DELAY and LOSS, we use tc \cite{tc} to manipulate the container traffic. For each combination of fault type and targeted service, we repeat the fault injections and data collections three times, resulting in a total of 90 collected failure cases for each benchmark microservice system. Table \ref{tab:real-data} presents the statistics summarizing the collected data.

\textbf{Blind spots in the benchmark systems.} Despite their widespread use for benchmarking, these systems have several blind spots (i.e., services without traces) by default. In the Online Boutique system, 7 out of 11 services are instrumented with tracing, leaving \textbf{4 blind spots}. The Sock Shop system is not instrumented at all, meaning \textbf{no traces are available} to construct a service call graph. In the Train Ticket system, 27 out of 64 services are instrumented with tracing, resulting in \textbf{37 blind spots}. Existing RCA methods that rely on call graphs~\cite{lee2023eadro, yu2023nezha, zhang2023diagfusion} will be unable to diagnose root causes in these blind spots, as they do not appear in the call graph.

\subsection{Evaluation Metrics}

Following existing works \cite{Azam2022rcd, pham2024baro, pham2024root},
we use two standard evaluation metrics: $AC@k$ and $Avg@k$ to measure the RCA performance. Given a set of failure cases A, $AC@k$ is calculated as follows,
\begin{equation}
    AC@k = \frac{1}{|A|} \sum\nolimits_{a\in A}\frac{\sum_{i<k}R^a[i]\in V^a_{rc}}{min(k, |V^a_{rc}|)},
\end{equation}
where $R^a[i]$ is the $i$th ranking result for the failure case $a$ by an RCA method, and $V^a_{rc}$ is the true root cause set of case $a$. $AC@k$ represents the probability the top $k$ results of the given method include the true root causes. Its values range from $0$ to $1$, with higher values indicating better performance. $Avg@k$, which shows the overall RCA performance, is measured as $Avg@k = \frac{1}{k}\sum_{j=1}^k AC@j$. Due to space constraints, we also refer to $AC@1$, $AC@3$, and $Avg@5$ as T1, T3, and A5, respectively.

We run all experiments on Linux servers each with 8 CPUs and 16GB RAM. In addition, we repeat each experiment five times and report the average results to minimize the impact of randomness. We use one-way ANOVA to assess overall differences among methods and pairwise t-tests for comparing individual method pairs. We report results as statistically significant when $p < 0.05$.

\subsection{Baselines} \label{sec:baselines}

We select nine RCA baselines from previous studies for performance comparison with our proposed multi-source RCA method, namely: PDiagnose \cite{hou2021pdiagnose}, HeMiRCA \cite{zhu2024hemirca}, CausalRCA \cite{Xin2023CausalRCA}, MicroCause \cite{Meng2020Microcause}, RCD \cite{Azam2022rcd}, CIRCA \cite{Li2022Circa}, BARO \cite{pham2024baro}, MicroRank \cite{yu2021microrank}, and TraceRCA \cite{dan2021tracerca}. Detailed information of these methods is as follows:

\begin{itemize}[leftmargin=*]
\item \textit{PDiagnose \cite{hou2021pdiagnose}:} PDiagnose is a multi-source RCA method that transforms metrics, logs, and traces into time series and determines root causes through voting. It relies on traces to determine the root cause services, and metrics to derive fine-grained root causes.

\item \textit{HeMiRCA \cite{zhu2024hemirca}}: HeMiRCA is a multi-source RCA method, which relies on the monotonic correlation between metrics and trace-based anomaly scores. HeMiRCA first measures trace-based anomaly scores and then exploits the correlations between metrics and the trace anomaly scores to rank the suspicious metrics and microservices. However, HeMiRCA does not use logs.

% \item \textit{MSCRED \cite{zhang2019deep}}: MSCRED takes multi-source time series data to perform anomaly detection and RCA. It relies on an encoder-decoder involving convolutional neural network with long-short term memory to reconstruct the correlation matrix to find the anomalies. The root cause of anomalies is the time series that has the most correlation changes.

% \item \textit{RUN \cite{run2024aaai}}: RUN takes multi-source time series data for RCA. It uses neural Granger causal discovery with contrastive learning to construct the causal graph among multi-source time series. Then, it applies PageRank with a personalized vector to recommend the top-k root causes.

\item \textit{CausalRCA \cite{Xin2023CausalRCA}:} CausalRCA constructs the causal graph from time series derived from metrics data using DAG-GNN \cite{Yu2019DagGNN}, a gradient-based causal structure learning method. Then it employs PageRank to rank the root causes from the estimated graph.

% \item \textit{PC-PageRank \cite{Xin2023CausalRCA}:} %Similar to CausalRCA, 
% PC-PageRank uses the PC algorithm \cite{Spirtes1995fci} to estimate the causal graph and then uses the PageRank algorithm to rank the root causes. Hereafter, we refer to it as PC-PR.

\item \textit{MicroCause \cite{Meng2020Microcause}:} MicroCause uses PCMCI \cite{Runge2019PCMCI} to construct the causal graph. %, akin to CausalRCA and PC-PR. 
Then, it applies temporal cause-oriented random walk to rank the root causes from the estimated causal graph.

\item \textit{RCD \cite{Azam2022rcd}:} RCD adopts a divide-and-conquer strategy to partition time series into chunks. Then, it uses the $\Psi$-PC  \cite{Jaber2020PsiFCI} to build causal graphs and identify root causes within each chunk. Recursively, it combines these root causes and iterates until only one chunk remains.

\item \textit{CIRCA \cite{Li2022Circa}:} CIRCA relies on a provided call graph to construct a causal graph. It then uses regression-based hypothesis testing analysis to identify root causes.

% \item \textit{$\epsilon$-Diagnosis \cite{Shan2019Ediagnosis}:} $\epsilon$-Diagnosis uses $\epsilon$-statistics and two-sample test algorithm to measure the similarity between every pair of time series and rank the root causes based on test scores. 

\item \textit{BARO \cite{pham2024baro}:} %Similar to $\epsilon$-Diagnosis, BARO is also a statistical analysis method, which 
BARO uses a nonparametric hypothesis testing technique based on median and IQR to measure the change of metrics time series after the failure time and rank the root causes.

\item \textit{MicroRank \cite{yu2021microrank}:} MicroRank is a trace-based RCA approach that combines personalized PageRank and Spectrum method to identify suspicious root causes from the collected trace data.

\item \textit{TraceRCA \cite{dan2021tracerca}:} TraceRCA uses spectrum analysis to identify the root cause services, based on the insight that a service with more abnormal and fewer normal traces passing through it is more likely to be the root cause.
\end{itemize}

For the baselines HeMiRCA, CausalRCA, MicroCause, RCD, CIRCA, BARO, MicroRank, and TraceRCA, we use their publicly available implementation and default hyperparameter settings suggested in their respective papers. We verified their correctness of the obtained source code  by reproducing the presented results in the original and related papers. For PDiagnose, we follow previous works \cite{yu2023nezha, zhang2023diagfusion, hou2021pdiagnose} to implement it since its source code is unavailable. Furthermore, recent multi-source RCA methods \cite{lee2023eadro, yu2023nezha, zhang2023diagfusion, zhang2021cloudrca, li2022actionable} exhibit limitations that prevent us from adopting them as baselines. Specifically,  some methods~\cite{lee2023eadro, zhang2023diagfusion, li2022actionable, zhang2021cloudrca} require labeled training data, which is unavailable to us, while~\cite{yu2023nezha} requires manual effort to integrate \texttt{trace\_id} into every log line. In contrast, our method does not require labeled data or heavy instrumentation into the systems.

It is important to note that four of our selected baselines (CausalRCA, MicroCause, RCD, BARO) are metric-based methods that do not require call graphs or trace data. They take time series as input and perform RCA using causal discovery or statistical analysis. These methods represent state-of-the-art metric-based RCA approaches in the literature.

\begin{table}[ht]
\centering
\caption{RCA performance of TORAI and baselines on the Online Boutique dataset, across six fault types. The best results are in \textbf{bold} iff the t-test reported a significant difference compared to other baselines ($p<0.05$).}
\label{tab:rq1-ob}
\vspace{-3pt}
\resizebox{\textwidth}{!}{ \setlength\tabcolsep{2pt}
\begin{tabular}{c|l|rrr|rrr|rrr|rrr|rrr|rrr|rrr}
\hline
\multirow{2}{*}{\begin{tabular}[c]{@{}c@{}}Data\\ Source\end{tabular}} & \multicolumn{1}{c|}{\multirow{2}{*}{Method}} & \multicolumn{3}{c|}{\textbf{CPU}} & \multicolumn{3}{c|}{\textbf{MEM}} & \multicolumn{3}{c|}{\textbf{DISK}} & \multicolumn{3}{c|}{\textbf{SOCKET}} & \multicolumn{3}{c|}{\textbf{DELAY}} & \multicolumn{3}{c|}{\textbf{LOSS}} & \multicolumn{3}{c}{\textbf{AVERAGE}} \\ \cline{3-23} 
 & \multicolumn{1}{c|}{} & \multicolumn{1}{c}{\textit{T1}} & \multicolumn{1}{c}{\textit{T3}} & \multicolumn{1}{c|}{\textit{A5}} & \multicolumn{1}{c}{\textit{T1}} & \multicolumn{1}{c}{\textit{T3}} & \multicolumn{1}{c|}{\textit{A5}} & \multicolumn{1}{c}{\textit{T1}} & \multicolumn{1}{c}{\textit{T3}} & \multicolumn{1}{c|}{\textit{A5}} & \multicolumn{1}{c}{\textit{T1}} & \multicolumn{1}{c}{\textit{T3}} & \multicolumn{1}{c|}{\textit{A5}} & \multicolumn{1}{c}{\textit{T1}} & \multicolumn{1}{c}{\textit{T3}} & \multicolumn{1}{c|}{\textit{A5}} & \multicolumn{1}{c}{\textit{T1}} & \multicolumn{1}{c}{\textit{T3}} & \multicolumn{1}{c|}{\textit{A5}} & \multicolumn{1}{c}{\textit{T1}} & \multicolumn{1}{c}{\textit{T3}} & \multicolumn{1}{c}{\textit{A5}} \\ \hline \hline
\multirow{6}{*}{Metric}
 & BARO & 0.47 & 0.80 & 0.72 & \textbf{0.93} & \textbf{1.00} & \textbf{0.99} & \textbf{1.00} & \textbf{1.00} & \textbf{1.00} & 0.60 & 0.87 & 0.83 & 0.47 & 0.67 & 0.63 & 0.53 & 0.60 & 0.64 & 0.67 & 0.82 & 0.80 \\ 
 & CausalRCA & 0.20 & 0.60 & 0.53 & 0.33 & 0.87 & 0.77 & 0.07 & 0.67 & 0.52 & 0.27 & 0.67 & 0.61 & 0.20 & 0.73 & 0.61 & 0.20 & 0.53 & 0.53 & 0.21 & 0.68 & 0.60 \\
 & CIRCA & 0.73 & 0.93 & 0.88 & 0.67 & \underline{0.93} & 0.89 & 0.80 & 0.87 & 0.85 & 0.67 & 0.87 & 0.83 & 0.47 & 0.80 & 0.75 & 0.67 & \underline{0.93} & 0.88 & 0.67 & 0.89 & 0.85 \\
 & MicroCause & 0.20 & 0.33 & 0.33 & 0.07 & 0.20 & 0.23 & 0.27 & 0.40 & 0.37 & 0.27 & 0.40 & 0.37 & 0.00 & 0.07 & 0.07 & 0.00 & 0.07 & 0.11 & 0.14 & 0.25 & 0.25 \\
 & RCD & \underline{0.87} & \textbf{1.00} & \underline{0.94} & 0.67 & 0.87 & 0.81 & 0.73 & 0.87 & 0.81 & 0.80 & \underline{0.93} & \underline{0.91} & 0.27 & 0.60 & 0.52 & 0.20 & 0.53 & 0.44 & 0.59 & 0.80 & 0.74 \\
 \hline
\multirow{6}{*}{Log}
  & BARO & 0.00 & 0.00 & 0.00 & 0.00 & 0.07 & 0.07 & 0.07 & 0.13 & 0.12 & 0.00 & 0.07 & 0.09 & 0.00 & 0.00 & 0.00 & 0.07 & 0.13 & 0.11 & 0.02 & 0.07 & 0.06 \\
  & CausalRCA & 0.00 & 0.13 & 0.15 & 0.07 & 0.13 & 0.16 & 0.00 & 0.00 & 0.12 & 0.00 & 0.13 & 0.13 & 0.00 & 0.27 & 0.25 & 0.07 & 0.67 & 0.52 & 0.02 & 0.22 & 0.22 \\
  & CIRCA & 0.00 & 0.20 & 0.24 & 0.07 & 0.27 & 0.27 & 0.13 & 0.27 & 0.32 & 0.13 & 0.27 & 0.32 & 0.00 & 0.07 & 0.13 & 0.07 & 0.13 & 0.19 & 0.07 & 0.20 & 0.25 \\
  & MicroCause & 0.27 & 0.67 & 0.57 & 0.13 & 0.40 & 0.39 & 0.31 & 0.69 & 0.68 & 0.13 & 0.47 & 0.47 & 0.00 & 0.07 & 0.11 & 0.33 & 0.47 & 0.51 & 0.20 & 0.46 & 0.46 \\
  & RCD & 0.00 & 0.07 & 0.05 & 0.00 & 0.07 & 0.05 & 0.00 & 0.07 & 0.04 & 0.00 & 0.13 & 0.11 & 0.07 & 0.13 & 0.12 & 0.27 & 0.33 & 0.32 & 0.06 & 0.13 & 0.12 \\

  \hline
\multirow{8}{*}{Trace}
 & BARO & 0.40 & 0.80 & 0.72 & 0.07 & 0.33 & 0.28 & 0.27 & 0.47 & 0.49 & 0.53 & 0.60 & 0.61 & 0.00 & 0.07 & 0.07 & 0.27 & 0.60 & 0.51 & 0.26 & 0.48 & 0.45 \\
 & CausalRCA & 0.30 & 0.37 & 0.36 & 0.23 & 0.33 & 0.36 & 0.07 & 0.20 & 0.19 & 0.20 & 0.27 & 0.27 & 0.00 & 0.03 & 0.03 & 0.03 & 0.30 & 0.30 & 0.14 & 0.25 & 0.25 \\
 & CIRCA & 0.40 & 0.60 & 0.55 & 0.27 & 0.53 & 0.53 & 0.33 & 0.67 & 0.59 & 0.33 & 0.53 & 0.52 & 0.07 & 0.13 & 0.12 & 0.73 & 0.80 & 0.83 & 0.36 & 0.54 & 0.52 \\
 & MicroCause & 0.17 & 0.33 & 0.27 & 0.00 & 0.00 & 0.04 & 0.10 & 0.40 & 0.40 & 0.00 & 0.20 & 0.32 & 0.00 & 0.00 & 0.00 & 0.00 & 0.00 & 0.00 & 0.05 & 0.16 & 0.17 \\
 & RCD & 0.47 & 0.49 & 0.49 & 0.31 & 0.38 & 0.38 & 0.17 & 0.43 & 0.37 & 0.24 & 0.52 & 0.46 & 0.00 & 0.04 & 0.03 & 0.04 & 0.06 & 0.06 & 0.21 & 0.32 & 0.30 \\
 & MicroRank & 0.00 & 0.40 & 0.36 & 0.00 & 0.40 & 0.36 & 0.00 & 0.40 & 0.36 & 0.00 & 0.40 & 0.36 & 0.00 & 0.40 & 0.36 & 0.00 & 0.27 & 0.19 & 0.00 & 0.38 & 0.33 \\
 & TraceRCA & 0.07 & 0.80 & 0.69 & 0.00 & 0.53 & 0.57 & 0.20 & 0.80 & 0.72 & 0.07 & 0.73 & 0.67 & 0.20 & 0.53 & 0.57 & 0.07 & 0.53 & 0.49 & 0.10 & 0.65 & 0.62 \\
 \hline
\multirow{7}{*}{\begin{tabular}[c]{@{}c@{}}Metric\\ +\\ Log\end{tabular}} 
  & BARO & 0.47 & 0.80 & 0.75 & \textbf{0.93} & \textbf{1.00} & \textbf{0.99} & \textbf{1.00} & \textbf{1.00} & \textbf{1.00} & 0.60 & 0.80 & 0.79 & 0.47 & 0.67 & 0.61 & 0.67 & 0.67 & 0.71 & 0.69 & 0.82 & 0.81 \\
 & CausalRCA & 0.13 & 0.30 & 0.30 & 0.12 & 0.24 & 0.25 & 0.00 & 0.07 & 0.10 & 0.00 & 0.13 & 0.15 & 0.00 & 0.30 & 0.28 & 0.07 & 0.50 & 0.41 & 0.09 & 0.24 & 0.22 \\
 & CIRCA & 0.00 & 0.07 & 0.04 & 0.00 & 0.00 & 0.00 & 0.00 & 0.00 & 0.00 & 0.00 & 0.00 & 0.00 & 0.00 & 0.00 & 0.00 & 0.00 & 0.07 & 0.07 & 0.00 & 0.02 & 0.02 \\
 & MicroCause & 0.29 & 0.50 & 0.46 & 0.07 & 0.33 & 0.31 & 0.21 & 0.43 & 0.43 & 0.27 & 0.67 & 0.61 & 0.07 & 0.20 & 0.19 & 0.13 & 0.50 & 0.48 & 0.17 & 0.44 & 0.41 \\
 & RCD & 0.79 & \underline{0.95} & 0.91 & 0.45 & 0.79 & 0.71 & 0.77 & 0.84 & 0.84 & \underline{0.91} & \underline{0.96} & \underline{0.95} & 0.36 & 0.67 & 0.60 & 0.29 & 0.67 & 0.59 & 0.60 & 0.81 & 0.77 \\
 \cline{2-23} 
 & \textbf{TORAI} & 0.67 & 0.80 & 0.81 & \underline{0.87} & \textbf{1.00} & \underline{0.97} & \underline{0.93} & \textbf{1.00} & \underline{0.95} & 0.80 & 0.87 & 0.84 & \textbf{0.73} & \textbf{1.00} & \textbf{0.92} & \underline{0.77} & \textbf{1.00} & \underline{0.93} & \underline{0.80} & \textbf{0.95} & \underline{0.91} \\ \hline
\multirow{8}{*}{\begin{tabular}[c]{@{}c@{}}Metric\\ +\\ Log\\ +\\ Trace\end{tabular}} 
 & BARO & 0.47 & 0.80 & 0.75 & \textbf{0.93} & \textbf{1.00} & \textbf{0.99} & \textbf{1.00} & \textbf{1.00} & \textbf{1.00} & 0.60 & 0.80 & 0.79 & 0.47 & 0.67 & 0.61 & 0.67 & 0.67 & 0.71 & 0.69 & 0.82 & 0.81 \\
 & CausalRCA & 0.20 & 0.27 & 0.27 & 0.13 & 0.27 & 0.29 & 0.07 & 0.07 & 0.09 & 0.07 & 0.20 & 0.23 & 0.07 & 0.33 & 0.31 & 0.07 & 0.27 & 0.28 & 0.10 & 0.24 & 0.25 \\
 & CIRCA & 0.00 & 0.07 & 0.04 & 0.00 & 0.00 & 0.00 & 0.00 & 0.00 & 0.00 & 0.00 & 0.00 & 0.00 & 0.00 & 0.00 & 0.00 & 0.00 & 0.00 & 0.03 & 0.00 & 0.01 & 0.01 \\
 & MicroCause & 0.50 & 0.67 & 0.60 & 0.25 & 0.25 & 0.25 & 0.33 & 0.78 & 0.76 & 0.50 & 0.50 & 0.50 & 0.00 & 0.33 & 0.30 & 0.14 & 0.29 & 0.37 & 0.29 & 0.47 & 0.46 \\
 & PDiagnose & 0.00 & 0.80 & 0.60 & 0.00 & 0.40 & 0.51 & 0.00 & 0.73 & 0.59 & 0.00 & 0.60 & 0.52 & 0.20 & 0.40 & 0.55 & 0.00 & 0.47 & 0.47 & 0.03 & 0.57 & 0.54 \\ 
 & HeMiRCA & 0.43 & 0.77 & 0.72 & 0.23 & 0.36 & 0.37 & 0.89 & \underline{0.95} & 0.94 & 0.20 & 0.40 & 0.43 & 0.33 & 0.53 & 0.55 & 0.40 & 0.40 & 0.40 & 0.41 & 0.57 & 0.57 \\
 & RCD & 0.80 & \textbf{1.00} & \underline{0.94} & 0.47 & 0.80 & 0.71 & 0.80 & \textbf{1.00} & \underline{0.95} & \textbf{1.00} & \textbf{1.00} & \textbf{1.00} & 0.33 & 0.40 & 0.51 & 0.40 & 0.60 & 0.60 & 0.63 & 0.80 & 0.79 \\
 \cline{2-23} 
 & \textbf{TORAI} & \textbf{0.92} & \textbf{1.00} & \textbf{0.98} & 0.67 & \textbf{1.00} & 0.93 & \textbf{1.00} & \textbf{1.00} & \textbf{1.00} & 0.87 & 0.93 & \underline{0.95} & \underline{0.67} & 0.73 & \underline{0.76} & \textbf{0.87} & \textbf{1.00} & \textbf{0.96} & \textbf{0.83} & \textbf{0.95} & \textbf{0.93} \\ \hline
\end{tabular} }

\vspace{0.1cm}

{\footnotesize (*) T1, T3, and A5 denote AC@1, AC@3, and Avg@5, respectively.}
\end{table}

\begin{table*}[ht]
\vspace{-3pt}
\caption{RCA performance of TORAI and baselines on the Sock Shop dataset, across six fault types. The best results are in \textbf{bold} iff the t-test reported a significant difference compared to other baselines ($p<0.05$).}
\vspace{-3pt}
\label{tab:rq1-ss}
\resizebox{\textwidth}{!}{ \setlength\tabcolsep{2pt}
\begin{tabular}{c|l|rrr|rrr|rrr|rrr|rrr|rrr|rrr}
\hline
\multirow{2}{*}{\begin{tabular}[c]{@{}c@{}}Data\\ Source\end{tabular}} & \multicolumn{1}{c|}{\multirow{2}{*}{Method}} & \multicolumn{3}{c|}{\textbf{CPU}} & \multicolumn{3}{c|}{\textbf{MEM}} & \multicolumn{3}{c|}{\textbf{DISK}} & \multicolumn{3}{c|}{\textbf{SOCKET}} & \multicolumn{3}{c|}{\textbf{DELAY}} & \multicolumn{3}{c|}{\textbf{LOSS}} & \multicolumn{3}{c}{\textbf{AVERAGE}} \\ \cline{3-23} 
 & \multicolumn{1}{c|}{} & \multicolumn{1}{c}{\textit{T1}} & \multicolumn{1}{c}{\textit{T3}} & \multicolumn{1}{c|}{\textit{A5}} & \multicolumn{1}{c}{\textit{T1}} & \multicolumn{1}{c}{\textit{T3}} & \multicolumn{1}{c|}{\textit{A5}} & \multicolumn{1}{c}{\textit{T1}} & \multicolumn{1}{c}{\textit{T3}} & \multicolumn{1}{c|}{\textit{A5}} & \multicolumn{1}{c}{\textit{T1}} & \multicolumn{1}{c}{\textit{T3}} & \multicolumn{1}{c|}{\textit{A5}} & \multicolumn{1}{c}{\textit{T1}} & \multicolumn{1}{c}{\textit{T3}} & \multicolumn{1}{c|}{\textit{A5}} & \multicolumn{1}{c}{\textit{T1}} & \multicolumn{1}{c}{\textit{T3}} & \multicolumn{1}{c|}{\textit{A5}} & \multicolumn{1}{c}{\textit{T1}} & \multicolumn{1}{c}{\textit{T3}} & \multicolumn{1}{c}{\textit{A5}} \\ \hline \hline
\multirow{5}{*}{Metric}
  & BARO & 0.00 & \textbf{1.00} & 0.80 & 0.20 & \textbf{1.00} & 0.83 & 0.00 & \underline{0.93} & 0.77 & 0.00 & \underline{0.93} & 0.71 & 0.00 & \underline{0.87} & 0.68 & 0.20 & \textbf{1.00} & 0.80 & 0.07 & \textbf{0.96} & 0.76 \\
 & CausalRCA & 0.20 & 0.60 & 0.55 & 0.40 & 0.80 & 0.75 & 0.20 & 0.60 & 0.55 & 0.33 & 0.60 & 0.60 & 0.27 & 0.40 & 0.43 & 0.00 & 0.33 & 0.32 & 0.23 & 0.56 & 0.53 \\
 & CIRCA & \underline{0.87} & \textbf{1.00} & \underline{0.97} & \underline{0.87} & \underline{0.93} & \underline{0.95} & \underline{0.87} & 0.87 & \underline{0.89} & \underline{0.67} & \textbf{1.00} & \underline{0.92} & \underline{0.67} & \underline{0.87} & \underline{0.85} & \underline{0.47} & 0.87 & \underline{0.81} & \underline{0.74} & \underline{0.92} & \underline{0.90} \\
 & MicroCause & 0.07 & 0.13 & 0.16 & 0.00 & 0.20 & 0.23 & 0.00 & 0.13 & 0.12 & 0.33 & 0.40 & 0.44 & 0.13 & 0.27 & 0.28 & 0.07 & 0.27 & 0.24 & 0.10 & 0.23 & 0.25 \\ 
 & RCD & 0.47 & 0.73 & 0.68 & 0.27 & 0.40 & 0.36 & 0.47 & 0.67 & 0.61 & 0.47 & 0.87 & 0.77 & 0.40 & 0.73 & 0.64 & 0.20 & 0.40 & 0.35 & 0.38 & 0.63 & 0.57 \\ \hline
\multirow{5}{*}{Log}
 & BARO & 0.20 & 0.47 & 0.51 & 0.20 & 0.47 & 0.48 & 0.13 & 0.33 & 0.39 & 0.13 & 0.33 & 0.39 & 0.20 & 0.33 & 0.40 & 0.13 & 0.60 & 0.52 & 0.17 & 0.42 & 0.45 \\
 & CIRCA & 0.13 & 0.53 & 0.48 & 0.00 & 0.33 & 0.29 & 0.07 & 0.47 & 0.41 & 0.00 & 0.40 & 0.36 & 0.07 & 0.33 & 0.36 & 0.20 & 0.47 & 0.52 & 0.08 & 0.42 & 0.40 \\
 & CausalRCA & 0.10 & 0.33 & 0.37 & 0.10 & 0.53 & 0.47 & 0.23 & 0.43 & 0.42 & 0.20 & 0.50 & 0.47 & 0.10 & 0.20 & 0.21 & \underline{0.47} & 0.80 & 0.76 & 0.20 & 0.47 & 0.45 \\ 
 & MicroCause & 0.50 & 0.75 & 0.70 & 0.29 & 0.71 & 0.60 & 0.00 & 0.33 & 0.27 & 0.13 & 0.38 & 0.38 & 0.00 & 0.17 & 0.13 & 0.00 & 0.43 & 0.34 & 0.15 & 0.46 & 0.40 \\ 
 & RCD & 0.07 & 0.21 & 0.18 & 0.09 & 0.16 & 0.14 & 0.12 & 0.33 & 0.28 & 0.09 & 0.19 & 0.17 & 0.08 & 0.15 & 0.13 & 0.37 & 0.48 & 0.46 & 0.14 & 0.25 & 0.23 \\ \hline
\multirow{8}{*}{\begin{tabular}[c]{@{}c@{}}Metric\\ +\\ Log\end{tabular}} 
  & BARO & 0.00 & \textbf{1.00} & 0.80 & 0.20 & \textbf{1.00} & 0.83 & 0.00 & \textbf{1.00} & 0.79 & 0.00 & \underline{0.93} & 0.71 & 0.00 & 0.80 & 0.65 & 0.20 & \textbf{1.00} & 0.80 & 0.07 & \textbf{0.96} & 0.76 \\
 & CausalRCA & 0.20 & 0.33 & 0.33 & 0.47 & 0.73 & 0.73 & 0.20 & 0.40 & 0.44 & 0.07 & 0.40 & 0.35 & 0.20 & 0.53 & 0.48 & 0.20 & 0.47 & 0.51 & 0.22 & 0.48 & 0.47 \\
 & CIRCA & 0.00 & 0.00 & 0.00 & 0.07 & 0.07 & 0.08 & 0.00 & 0.00 & 0.05 & 0.00 & 0.00 & 0.00 & 0.00 & 0.07 & 0.07 & 0.00 & 0.20 & 0.19 & 0.01 & 0.06 & 0.07 \\
 & MicroCause & 0.33 & 0.33 & 0.40 & 0.17 & 0.33 & 0.37 & 0.00 & 0.67 & 0.53 & 0.25 & 0.75 & 0.63 & 0.20 & 0.80 & 0.72 & 0.00 & 0.40 & 0.36 & 0.16 & 0.55 & 0.50 \\
 & RCD & 0.61 & \underline{0.77} & 0.74 & 0.20 & 0.49 & 0.42 & 0.36 & 0.64 & 0.59 & \underline{0.67} & 0.89 & 0.85 & 0.28 & 0.67 & 0.58 & 0.19 & \underline{0.92} & 0.76 & 0.39 & 0.73 & 0.66 \\
 \cline{2-23} 
 & \textbf{TORAI} & \textbf{0.93} & \textbf{1.00} & \textbf{0.98} & \textbf{1.00} & \textbf{1.00} & \textbf{1.00} & \textbf{0.93} & \underline{0.93} & \textbf{0.95} & \textbf{0.84} & \underline{0.93} & \textbf{0.94} & \textbf{0.75} & \textbf{0.88} & \textbf{0.86} & \textbf{0.60} & \textbf{1.00} & \textbf{0.92} & \textbf{0.84} & \textbf{0.96} & \textbf{0.94} \\ \hline
\end{tabular} }

\vspace{0.1cm}

{\footnotesize (*) T1, T3, and A5 denote AC@1, AC@3, and Avg@5, respectively.}
\end{table*}

\begin{table}[ht]
\centering
\caption{RCA performance of TORAI and baselines on the Train Ticket dataset, across six fault types. The best results are in \textbf{bold} iff the t-test reported a significant difference compared to other baselines ($p<0.05$).}
\label{tab:rq1-tt}
\resizebox{\textwidth}{!}{ \setlength\tabcolsep{2pt}
\begin{tabular}{c|l|rrr|rrr|rrr|rrr|rrr|rrr|rrr}
\hline
\multirow{2}{*}{\begin{tabular}[c]{@{}c@{}}Data\\ Source\end{tabular}} & \multicolumn{1}{c|}{\multirow{2}{*}{Method}} & \multicolumn{3}{c|}{\textbf{CPU}} & \multicolumn{3}{c|}{\textbf{MEM}} & \multicolumn{3}{c|}{\textbf{DISK}} & \multicolumn{3}{c|}{\textbf{SOCKET}} & \multicolumn{3}{c|}{\textbf{DELAY}} & \multicolumn{3}{c|}{\textbf{LOSS}} & \multicolumn{3}{c}{\textbf{AVERAGE}} \\ \cline{3-23} 
 & \multicolumn{1}{c|}{} & \multicolumn{1}{c}{\textit{T1}} & \multicolumn{1}{c}{\textit{T3}} & \multicolumn{1}{c|}{\textit{A5}} & \multicolumn{1}{c}{\textit{T1}} & \multicolumn{1}{c}{\textit{T3}} & \multicolumn{1}{c|}{\textit{A5}} & \multicolumn{1}{c}{\textit{T1}} & \multicolumn{1}{c}{\textit{T3}} & \multicolumn{1}{c|}{\textit{A5}} & \multicolumn{1}{c}{\textit{T1}} & \multicolumn{1}{c}{\textit{T3}} & \multicolumn{1}{c|}{\textit{A5}} & \multicolumn{1}{c}{\textit{T1}} & \multicolumn{1}{c}{\textit{T3}} & \multicolumn{1}{c|}{\textit{A5}} & \multicolumn{1}{c}{\textit{T1}} & \multicolumn{1}{c}{\textit{T3}} & \multicolumn{1}{c|}{\textit{A5}} & \multicolumn{1}{c}{\textit{T1}} & \multicolumn{1}{c}{\textit{T3}} & \multicolumn{1}{c}{\textit{A5}} \\ \hline
\multirow{4}{*}{Metric} 
 & BARO & 0.47 & 0.80 & 0.72 & \underline{0.93} & \textbf{1.00} & \underline{0.99} & \textbf{1.00} & \textbf{1.00} & \textbf{1.00} & 0.60 & \underline{0.87} & 0.83 & 0.47 & 0.67 & 0.63 & 0.53 & 0.60 & 0.64 & 0.67 & 0.82 & 0.80 \\
 & CausalRCA & 0.40 & 0.63 & 0.59 & 0.10 & 0.27 & 0.24 & 0.43 & \underline{0.83} & \underline{0.75} & 0.23 & 0.50 & 0.45 & 0.13 & 0.23 & 0.21 & 0.03 & 0.37 & 0.33 & 0.22 & 0.47 & 0.43 \\
 & CIRCA & 0.27 & 0.27 & 0.28 & 0.47 & 0.73 & 0.68 & 0.53 & 0.67 & 0.64 & 0.27 & 0.53 & 0.52 & 0.20 & 0.27 & 0.28 & 0.20 & 0.33 & 0.35 & 0.32 & 0.47 & 0.46 \\
 & MicroCause & 0.19 & 0.44 & 0.40 & 0.00 & 0.09 & 0.07 & 0.40 & 0.40 & 0.40 & 0.00 & 0.17 & 0.15 & 0.00 & 0.22 & 0.13 & 0.00 & 0.00 & 0.07 & 0.10 & 0.22 & 0.20 \\ 
 & RCD & 0.13 & 0.13 & 0.16 & 0.07 & 0.07 & 0.07 & 0.00 & 0.07 & 0.05 & 0.13 & 0.33 & 0.29 & 0.13 & 0.13 & 0.15 & 0.07 & 0.07 & 0.07 & 0.09 & 0.13 & 0.13 \\ \hline
\multirow{5}{*}{Log} 
 & BARO & 0.00 & 0.00 & 0.00 & 0.00 & 0.07 & 0.07 & 0.07 & 0.13 & 0.12 & 0.00 & 0.07 & 0.09 & 0.00 & 0.00 & 0.00 & 0.07 & 0.13 & 0.11 & 0.02 & 0.07 & 0.06 \\
 & CausalRCA & 0.07 & 0.20 & 0.22 & 0.07 & 0.10 & 0.09 & 0.07 & 0.10 & 0.12 & 0.00 & 0.07 & 0.08 & 0.03 & 0.20 & 0.15 & 0.10 & 0.20 & 0.20 & 0.06 & 0.15 & 0.14 \\
 & CIRCA & 0.13 & 0.20 & 0.27 & 0.07 & 0.33 & 0.27 & 0.07 & 0.07 & 0.12 & 0.07 & 0.20 & 0.23 & 0.13 & 0.33 & 0.33 & 0.13 & 0.33 & 0.33 & 0.10 & 0.24 & 0.26 \\
 & MicroCause & 0.00 & 0.00 & 0.10 & 0.20 & 0.40 & 0.40 & 0.14 & 0.14 & 0.14 & 0.13 & 0.38 & 0.28 & 0.11 & 0.11 & 0.16 & 0.11 & 0.22 & 0.20 & 0.12 & 0.21 & 0.21 \\ 
 & RCD & 0.07 & 0.16 & 0.14 & 0.12 & 0.15 & 0.14 & 0.07 & 0.15 & 0.13 & 0.06 & 0.19 & 0.17 & 0.11 & 0.24 & 0.21 & 0.12 & 0.16 & 0.15 & 0.09 & 0.18 & 0.16 \\ \hline
\multirow{8}{*}{Trace}
 & BARO & 0.40 & 0.80 & 0.72 & 0.07 & 0.33 & 0.28 & 0.27 & 0.47 & 0.49 & 0.53 & 0.60 & 0.61 & 0.00 & 0.07 & 0.07 & 0.27 & 0.60 & 0.51 & 0.26 & 0.48 & 0.45 \\
 & CausalRCA & 0.07 & 0.33 & 0.29 & 0.07 & 0.20 & 0.19 & 0.20 & 0.27 & 0.25 & 0.00 & 0.00 & 0.03 & 0.00 & 0.07 & 0.05 & 0.27 & 0.47 & 0.44 & 0.10 & 0.22 & 0.21 \\
 & CIRCA & 0.13 & 0.20 & 0.23 & 0.20 & 0.27 & 0.25 & 0.20 & 0.20 & 0.21 & 0.13 & 0.13 & 0.15 & 0.20 & 0.33 & 0.35 & 0.13 & 0.27 & 0.27 & 0.17 & 0.23 & 0.24 \\
 & MicroCause & 0.00 & 0.00 & 0.00 & 0.00 & 0.00 & 0.00 & 0.00 & 0.00 & 0.00 & 0.00 & 0.00 & 0.00 & 0.00 & 0.00 & 0.00 & 0.00 & 0.00 & 0.00 & 0.00 & 0.00 & 0.00 \\
 & MicroRank & 0.21 & 0.43 & 0.34 & 0.25 & 0.38 & 0.33 & 0.00 & 0.36 & 0.27 & 0.30 & 0.40 & 0.36 & 0.08 & 0.31 & 0.23 & 0.14 & 0.36 & 0.30 & 0.16 & 0.37 & 0.31 \\
 & RCD & 0.53 & \textbf{0.87} & 0.79 & 0.53 & 0.67 & 0.63 & \underline{0.67} & 0.67 & 0.69 & \underline{0.73} & 0.80 & 0.79 & 0.13 & 0.27 & 0.21 & 0.60 & 0.73 & \underline{0.71} & 0.53 & 0.67 & 0.64 \\
 & TraceRCA & \textbf{0.64} & 0.79 & 0.74 & 0.63 & \underline{0.88} & 0.83 & 0.64 & 0.71 & 0.74 & 0.60 & 0.80 & 0.76 & \underline{0.85} & \underline{0.85} & \textbf{0.88} & 0.57 & 0.71 & 0.67 & 0.66 & 0.79 & 0.77 \\ \hline
\multirow{5}{*}{\begin{tabular}[c]{@{}c@{}}Metric\\ +\\ Log\end{tabular}}
 & BARO & 0.47 & 0.80 & 0.75 & \underline{0.93} & \textbf{1.00} & \underline{0.99} & \textbf{1.00} & \textbf{1.00} & \textbf{1.00} & 0.60 & 0.80 & 0.79 & 0.47 & 0.67 & 0.61 & \underline{0.67} & 0.67 & \underline{0.71} & 0.69 & 0.82 & 0.81 \\
 & CIRCA & 0.00 & 0.13 & 0.09 & 0.00 & 0.07 & 0.08 & 0.07 & 0.07 & 0.09 & 0.00 & 0.13 & 0.15 & 0.00 & 0.07 & 0.05 & 0.07 & 0.07 & 0.09 & 0.02 & 0.09 & 0.09 \\
 & MicroCause & 0.00 & 0.00 & 0.00 & 0.07 & 0.07 & 0.07 & 0.00 & 0.00 & 0.00 & 0.07 & 0.07 & 0.08 & 0.00 & 0.00 & 0.00 & 0.07 & 0.13 & 0.13 & 0.04 & 0.05 & 0.05 \\ 
 & RCD & 0.13 & 0.27 & 0.23 & 0.00 & 0.07 & 0.08 & 0.13 & 0.27 & 0.21 & 0.20 & 0.33 & 0.29 & 0.00 & 0.13 & 0.12 & 0.07 & 0.13 & 0.11 & 0.09 & 0.20 & 0.17 \\
 \cline{2-23} 
 & \textbf{TORAI} & 0.53 & \textbf{0.87} & 0.80 & \textbf{1.00} & \textbf{1.00} & \textbf{1.00} & \textbf{1.00} & \textbf{1.00} & \textbf{1.00} & \textbf{0.80} & \underline{0.87} & \underline{0.87} & 0.53 & 0.67 & 0.64 & 0.57 & 0.63 & 0.67 & \underline{0.74} & \underline{0.84} & \underline{0.83} \\ \hline
\multirow{6}{*}{\begin{tabular}[c]{@{}c@{}}Metric\\ +\\ Log\\ +\\ Trace\end{tabular}} 
 & BARO & 0.47 & 0.80 & 0.75 & \underline{0.93} & \textbf{1.00} & \underline{0.99} & \textbf{1.00} & \textbf{1.00} & \textbf{1.00} & 0.60 & 0.80 & 0.79 & 0.47 & 0.67 & 0.61 & \underline{0.67} & 0.67 & \underline{0.71} & 0.69 & 0.82 & 0.81 \\
 & CIRCA & 0.00 & 0.07 & 0.09 & 0.07 & 0.13 & 0.21 & 0.00 & 0.07 & 0.09 & 0.07 & 0.13 & 0.16 & 0.07 & 0.07 & 0.07 & 0.13 & 0.20 & 0.17 & 0.06 & 0.11 & 0.13 \\
 & HeMiRCA & 0.00 & 0.07 & 0.15 & 0.13 & 0.27 & 0.27 & 0.13 & 0.20 & 0.19 & 0.20 & 0.47 & 0.43 & 0.07 & 0.33 & 0.32 & 0.07 & 0.13 & 0.15 & 0.10 & 0.25 & 0.25 \\
 & MicroCause & 0.07 & 0.13 & 0.11 & 0.07 & 0.07 & 0.07 & 0.00 & 0.07 & 0.11 & 0.00 & 0.00 & 0.03 & 0.00 & 0.00 & 0.00 & 0.00 & 0.13 & 0.16 & 0.02 & 0.07 & 0.08 \\
 & PDiagnose & \underline{0.60} & \textbf{0.87} & \underline{0.81} & 0.40 & 0.47 & 0.48 & 0.33 & 0.73 & 0.69 & 0.33 & 0.67 & 0.60 & \textbf{0.87} & \textbf{0.87} & \underline{0.87} & 0.33 & 0.60 & 0.57 & 0.48 & 0.70 & 0.67 \\ 
 & RCD & 0.17 & \underline{0.84} & 0.72 & 0.00 & 0.44 & 0.39 & 0.07 & 0.76 & 0.62 & 0.21 & 0.77 & 0.66 & 0.05 & 0.28 & 0.25 & 0.07 & \underline{0.75} & 0.60 & 0.10 & 0.64 & 0.54 \\
 \cline{2-23} 
 & \textbf{TORAI} & \underline{0.60} & \textbf{0.87} & \textbf{0.83} & \textbf{1.00} & \textbf{1.00} & \textbf{1.00} & \textbf{1.00} & \textbf{1.00} & \textbf{1.00} & \underline{0.73} & \textbf{1.00} & \textbf{0.93} & 0.56 & 0.67 & 0.66 & \textbf{0.73} & \textbf{1.00} & \textbf{0.92} & \textbf{0.77} & \textbf{0.92} & \textbf{0.89} \\ \hline
\end{tabular} }
% \vspace{10pt}

{\footnotesize (*) CausalRCA exceeds the limit of 2 hours per case. T1, T3, and A5 denote AC@1, AC@3, and Avg@5, respectively.}
\end{table}

\subsection{RQ1. Effectiveness in Coarse-grained Root Cause Analysis} \label{sec:experimental-rq1}

In this section, we evaluate the coarse-grained RCA performance of our proposed TORAI and the RCA baseline methods on all three datasets. 
Tables \ref{tab:rq1-ob}, \ref{tab:rq1-ss}, and \ref{tab:rq1-tt} report the overall performance of all methods on the Online Boutique, Sock Shop, and Train Ticket datasets at coarse-grained level, respectively.  
We calculate the accuracy for each type of fault: CPU hog (CPU), memory leak (MEM), disk IO stress (DISK), socket stress (SOCKET), network delay (DELAY), and packet loss (LOSS). Additionally, we report the AVERAGE scores to present the overall performance across fault types and data sources. We perform statistical analysis on the results using the t-tests to check the pairwise differences among all RCA methods. We \textbf{bold} the best result iff the statistical tests report a significant difference ($p<0.05$) compared to others. In the tables, the first column indicates the data sources used for the methods in the second column. We draw the following observations:

\textbf{(1) TORAI performs the best on all three datasets.} For example, on the Online Boutique dataset, TORAI achieves an average T1, T3, and A5 score of 0.83, 0.95, and 0.93 when diagnosing root cause service, while metric-based CausalRCA achieves 0.21, 0.68, and 0.6, respectively. On the Sock Shop dataset, TORAI achieves the averages of 0.84, 0.96, and 0.94 for T1, T3, and A5, respectively, while the multi-source version of RCD reaches 0.39, 0.73, and 0.66. On the Train Ticket dataset, the best average A5 scores of CausalRCA, RCD, and CIRCA are 0.43, 0.64, and 0.46, respectively. Our TORAI beats them with a large margin, achieving an A5 of 0.89.

\textbf{(2) TORAI effectively integrates multi-source data, surpassing most baselines.} For example, on the Train Ticket dataset, when incorporating metrics, logs, and traces, TORAI achieves the T1, T3, and A5 scores of 0.77, 0.92, and 0.89, respectively. In contrast, other RCA methods like CausalRCA, RCD, CIRCA, and MicroCause perform worse because their designs are less effective. These methods typically attempt to build causal graphs from all time series data and rely on scoring techniques like PageRank, random walk, or hypothesis testing, without leveraging clustering or severity scoring as TORAI does.

\textbf{(3) TORAI can diagnose failures in blind spots.} PDiagnose and HeMiRCA are multi-source RCA methods but they rely heavily on trace data to construct the service call graph. Consequently, they cannot perform RCA for the Sock Shop dataset, where there is no trace data to construct the call graph (i.e., all services are blind spots). Meanwhile, TORAI can still perform RCA effectively in the presence of blind spots using metrics and logs without requiring traces to construct a call graph (see Table~\ref{tab:rq1-ss}).
It is worth noting that metrics and logs are easier to collect as developers do not need to spend much effort to obtain them. Metrics can be automatically obtained by a monitoring system, and logs are naturally produced by the systems for troubleshooting purposes. In contrast, developers need to spend tremendous effort to instrument the system with distributed tracing \cite{shen2023deepflow}. 

\textbf{(4) On the large Train Ticket system, TORAI outperforms other methods by a significant margin.} For example, our method achieves an Avg@5 score of 0.89, while CausalRCA, CIRCA, and RCD achieve their best scores of 0.43, 0.46, and 0.64, respectively.  Across all faults, TORAI achieves Avg@5 scores of 0.83, 1, 1, 0.93, 0.66, and 0.92 for CPU, MEM, DISK, SOCKET, DELAY, and LOSS, respectively. This demonstrates the effectiveness of our proposed TORAI, which works not only on small demo systems but also on large systems with many services.

\subsection{RQ2. Effectiveness in Fine-grained Root Cause Analysis}

\begin{table}[t]
\centering
\caption{Fine-grained RCA performance of TORAI and baselines on the Online Boutique dataset, across six fault types. The best results are in \textbf{bold} iff the t-test reported a significant difference compared to other baselines ($p<0.05$). For all baselines, we select their best setup when taking different data sources.}
\label{tab:rq1-ob-fine}
\vspace{-10pt}
\resizebox{\textwidth}{!}{ \setlength\tabcolsep{3pt}
\begin{tabular}{l|rrr|rrr|rrr|rrr|rrr|rrr|rrr}
\hline
\multicolumn{1}{c|}{\multirow{2}{*}{Method}} & \multicolumn{3}{c|}{\textbf{CPU}} & \multicolumn{3}{c|}{\textbf{MEM}} & \multicolumn{3}{c|}{\textbf{DISK}} & \multicolumn{3}{c|}{\textbf{SOCKET}} & \multicolumn{3}{c|}{\textbf{DELAY}} & \multicolumn{3}{c|}{\textbf{LOSS}} & \multicolumn{3}{c}{\textbf{AVERAGE}} \\ \cline{2-22} 
\multicolumn{1}{c|}{} & \multicolumn{1}{c}{\textit{T1}} & \multicolumn{1}{c}{\textit{T3}} & \multicolumn{1}{c|}{\textit{A5}} & \multicolumn{1}{c}{\textit{T1}} & \multicolumn{1}{c}{\textit{T3}} & \multicolumn{1}{c|}{\textit{A5}} & \multicolumn{1}{c}{\textit{T1}} & \multicolumn{1}{c}{\textit{T3}} & \multicolumn{1}{c|}{\textit{A5}} & \multicolumn{1}{c}{\textit{T1}} & \multicolumn{1}{c}{\textit{T3}} & \multicolumn{1}{c|}{\textit{A5}} & \multicolumn{1}{c}{\textit{T1}} & \multicolumn{1}{c}{\textit{T3}} & \multicolumn{1}{c|}{\textit{A5}} & \multicolumn{1}{c}{\textit{T1}} & \multicolumn{1}{c}{\textit{T3}} & \multicolumn{1}{c|}{\textit{A5}} & \multicolumn{1}{c}{\textit{T1}} & \multicolumn{1}{c}{\textit{T3}} & \multicolumn{1}{c}{\textit{A5}} \\ \hline
BARO & 0 & 0.33 & 0.43 & 0.2 & \textbf{0.8} & 0.67 & 0 & 0 & 0 & 0 & 0 & 0 & 0 & \textbf{0.73} & 0.6 & 0.33 & 0.73 & 0.65 & 0.09 & 0.43 & 0.39 \\ 
CausalRCA & \textbf{0.2} & 0.53 & 0.45 & \textbf{0.47} & 0.67 & 0.65 & 0.07 & 0.27 & 0.32 & 0 & 0.4 & \textbf{0.35} & 0.27 & 0.67 & 0.61 & 0.13 & 0.27 & 0.23 & 0.19 & 0.47 & 0.44 \\
CIRCA & 0.13 & 0.47 & 0.44 & \textbf{0.47} & 0.67 & 0.64 & 0.87 & 0.87 & 0.87 & 0 & 0.07 & 0.08 & 0.47 & \textbf{0.73} & 0.67 & 0.33 & 0.6 & 0.61 & 0.38 & 0.57 & 0.55 \\
MicroCause & 0.13 & 0.4 & 0.32 & 0 & 0 & 0.04 & 0.04 & 0.31 & 0.29 & 0 & 0.07 & 0.09 & 0 & 0 & 0 & 0 & 0 & 0 & 0.03 & 0.13 & 0.12 \\
HeMiRCA & 0.25 & 0.27 & 0.29 & 0.1 & 0.15 & 0.15 & 0.77 & 0.82 & 0.82 & 0.07 & 0.07 & 0.1 & 0.13 & 0.23 & 0.25 & 0.4 & 0.4 & 0.4 & 0.29 & 0.32 & 0.34 \\
RCD & 0.07 & 0.07 & 0.07 & 0.27 & 0.33 & 0.31 & 0 & 0 & 0 & 0 & 0 & 0 & 0.33 & 0.33 & 0.4 & 0.2 & 0.2 & 0.2 & 0.15 & 0.16 & 0.16 \\
\hline
TORAI & 0.13 & \textbf{0.73} & \textbf{0.63} & 0.4 & 0.67 & 0.6 & \textbf{1} & \textbf{1} & \textbf{1} & 0 & 0.07 & \textbf{0.35} & \textbf{0.6} & 0.67 & 0.65 & \textbf{0.8} & \textbf{0.87} & \textbf{0.85} & \textbf{0.49} & \textbf{0.67} & \textbf{0.68} \\ \hline
\end{tabular} }
\vspace{-10pt}
\end{table}

\begin{table}[t]
\caption{Fine-grained RCA performance of TORAI and baselines on the Sock Shop dataset, across six fault types. The best results are in \textbf{bold} iff the t-test reported a significant difference compared to others ($p<0.05$). For all baselines, we select their best setup when taking different data sources.} \label{tab:rq1-ss-fine}
\vspace{-10pt}
\resizebox{\columnwidth}{!}{ \setlength\tabcolsep{3pt}
\begin{tabular}{l|rrr|rrr|rrr|rrr|rrr|rrr|rrr}
\hline
\multicolumn{1}{c|}{\multirow{2}{*}{Method}} & \multicolumn{3}{c|}{\textbf{CPU}} & \multicolumn{3}{c|}{\textbf{MEM}} & \multicolumn{3}{c|}{\textbf{DISK}} & \multicolumn{3}{c|}{\textbf{SOCKET}} & \multicolumn{3}{c|}{\textbf{DELAY}} & \multicolumn{3}{c|}{\textbf{LOSS}} & \multicolumn{3}{c}{\textbf{AVERAGE}} \\ \cline{2-22} 
\multicolumn{1}{c|}{} & \multicolumn{1}{c}{\textit{T1}} & \multicolumn{1}{c}{\textit{T3}} & \multicolumn{1}{c|}{\textit{A5}} & \multicolumn{1}{c}{\textit{T1}} & \multicolumn{1}{c}{\textit{T3}} & \multicolumn{1}{c|}{\textit{A5}} & \multicolumn{1}{c}{\textit{T1}} & \multicolumn{1}{c}{\textit{T3}} & \multicolumn{1}{c|}{\textit{A5}} & \multicolumn{1}{c}{\textit{T1}} & \multicolumn{1}{c}{\textit{T3}} & \multicolumn{1}{c|}{\textit{A5}} & \multicolumn{1}{c}{\textit{T1}} & \multicolumn{1}{c}{\textit{T3}} & \multicolumn{1}{c|}{\textit{A5}} & \multicolumn{1}{c}{\textit{T1}} & \multicolumn{1}{c}{\textit{T3}} & \multicolumn{1}{c|}{\textit{A5}} & \multicolumn{1}{c}{\textit{T1}} & \multicolumn{1}{c}{\textit{T3}} & \multicolumn{1}{c}{\textit{A5}} \\ \hline
BARO & 0 & 0.87 & 0.68 & 0.2 & \textbf{1} & 0.8 & 0 & 0 & 0 & 0 & 0 & 0.01 & 0 & \textbf{0.87} & 0.68 & 0.2 & \textbf{1} & 0.8 & 0.07 & 0.62 & 0.50 \\
CausalRCA & 0.27 & 0.63 & 0.58 & 0.3 & 0.67 & 0.59 & 0.07 & 0.17 & 0.17 & 0 & 0 & 0.03 & 0.33 & 0.57 & 0.56 & 0.23 & 0.37 & 0.37 & 0.2 & 0.37 & 0.36 \\
CIRCA & 0.4 & 0.8 & 0.72 & 0.8 & \textbf{1} & 0.95 & \textbf{0.67} & \textbf{0.67} & \textbf{0.67} & 0 & 0.2 & 0.16 & 0.53 & 0.8 & \textbf{0.79} & 0.6 & 0.87 & \textbf{0.85} & 0.5 & \textbf{0.69} & \textbf{0.67} \\
MicroCause & 0 & 0.09 & 0.07 & 0 & 0.08 & 0.05 & 0.09 & 0.18 & 0.18 & 0.07 & 0.2 & 0.23 & 0 & 0 & 0 & 0 & 0 & 0 & 0.03 & 0.08 & 0.08 \\ 
RCD & 0.11 & 0.15 & 0.15 & 0.05 & 0.16 & 0.15 & 0 & 0 & 0 & 0.13 & 0.19 & 0.2 & 0.24 & 0.27 & 0.27 & 0.03 & 0.03 & 0.03 & 0.09 & 0.13 & 0.13 \\
\hline
\textbf{TORAI} & \textbf{0.6} & \textbf{0.93} & \textbf{0.85} & \textbf{0.93} & \textbf{1} & \textbf{0.99} & \textbf{0.67} & \textbf{0.67} & \textbf{0.67} & 0 & 0.13 & \textbf{0.32} & \textbf{0.73} & 0.73 & 0.73 & 0.6 & 0.6 & 0.6 & \textbf{0.59} & \textbf{0.68} & \textbf{0.69} \\ \hline
\end{tabular} }
\vspace{-10pt}
\end{table}

\begin{table}[t]
\caption{Fine-grained RCA performance of TORAI and baselines on the Train Ticket dataset, across six fault types. The best results are in \textbf{bold} iff the t-test reported a significant difference compared to other baselines ($p<0.05$). For all baselines, we select their best setup when taking different data sources.} \label{tab:rq1-tt-fine}
\vspace{-10pt}
\resizebox{\columnwidth}{!}{ \setlength\tabcolsep{3pt}
\begin{tabular}{l|rrr|rrr|rrr|rrr|rrr|rrr|rrr}
\hline
\multicolumn{1}{c|}{\multirow{2}{*}{Method}} & \multicolumn{3}{c|}{\textbf{CPU}} & \multicolumn{3}{c|}{\textbf{MEM}} & \multicolumn{3}{c|}{\textbf{DISK}} & \multicolumn{3}{c|}{\textbf{SOCKET}} & \multicolumn{3}{c|}{\textbf{DELAY}} & \multicolumn{3}{c|}{\textbf{LOSS}} & \multicolumn{3}{c}{\textbf{AVERAGE}} \\ \cline{2-22} 
\multicolumn{1}{c|}{} & \multicolumn{1}{c}{\textit{T1}} & \multicolumn{1}{c}{\textit{T3}} & \multicolumn{1}{c|}{\textit{A5}} & \multicolumn{1}{c}{\textit{T1}} & \multicolumn{1}{c}{\textit{T3}} & \multicolumn{1}{c|}{\textit{A5}} & \multicolumn{1}{c}{\textit{T1}} & \multicolumn{1}{c}{\textit{T3}} & \multicolumn{1}{c|}{\textit{A5}} & \multicolumn{1}{c}{\textit{T1}} & \multicolumn{1}{c}{\textit{T3}} & \multicolumn{1}{c|}{\textit{A5}} & \multicolumn{1}{c}{\textit{T1}} & \multicolumn{1}{c}{\textit{T3}} & \multicolumn{1}{c|}{\textit{A5}} & \multicolumn{1}{c}{\textit{T1}} & \multicolumn{1}{c}{\textit{T3}} & \multicolumn{1}{c|}{\textit{A5}} & \multicolumn{1}{c}{\textit{T1}} & \multicolumn{1}{c}{\textit{T3}} & \multicolumn{1}{c}{\textit{A5}} \\ \hline
BARO & 0.07 & 0.33 & 0.32 & \textbf{0.33} & 0.8 & 0.75 & \textbf{1} & \textbf{1} & \textbf{1} & 0 & 0 & 0.01 & 0.47 & \textbf{0.67} & \textbf{0.63} & 0.53 & 0.53 & 0.53 & 0.4 & 0.56 & 0.54 \\
CausalRCA & \textbf{0.33} & 0.47 & \textbf{0.49} & 0 & 0.03 & 0.02 & 0.43 & 0.73 & 0.68 & 0 & 0 & 0 & 0.13 & 0.13 & 0.13 & 0.03 & 0.27 & 0.22 & 0.15 & 0.27 & 0.26 \\
CIRCA & 0.01 & 0.05 & 0.05 & 0 & 0.23 & 0.15 & 0.05 & 0.16 & 0.12 & 0.02 & 0.12 & 0.07 & 0 & 0 & 0.09 & 0 & 0.09 & 0.11 & 0.01 & 0.11 & 0.10 \\
HeMiRCA & 0 & 0.07 & 0.05 & 0.07 & 0.07 & 0.07 & 0 & 0 & 0 & 0 & 0 & 0 & 0.07 & 0.13 & 0.12 & 0 & 0 & 0 & 0.02 & 0.05 & 0.04 \\
MicroCause & 0.06 & 0.11 & 0.11 & 0 & 0.02 & 0.02 & 0.12 & 0.12 & 0.12 & 0 & 0.01 & 0.04 & 0 & 0 & 0.03 & 0 & 0 & 0 & 0.03 & 0.04 & 0.05 \\ 
RCD & 0.03 & 0.07 & 0.05 & 0 & 0 & 0 & 0 & 0 & 0 & 0.2 & 0.23 & 0.23 & 0 & 0 & 0 & 0 & 0 & 0 & 0.04 & 0.05 & 0.05 \\
\hline
\textbf{TORAI} & 0.2 & \textbf{0.6} & \textbf{0.51} & 0.27 & \textbf{1} & \textbf{0.85} & \textbf{1} & \textbf{1} & \textbf{1} & 0.07 & \textbf{0.47} & \textbf{0.43} & \textbf{0.6} & 0.6 & 0.6 & \textbf{0.6} & \textbf{0.73} & \textbf{0.68} & \textbf{0.46} & \textbf{0.73} & \textbf{0.68} \\ \hline
\end{tabular} }
\vspace{-10pt}
\end{table}

In this section, we evaluate the fine-grained RCA performance of our proposed TORAI alongside baseline RCA methods across all three datasets. Tables \ref{tab:rq1-ob-fine}, \ref{tab:rq1-ss-fine}, and \ref{tab:rq1-tt-fine} report the fine-grained RCA performance of all methods on the Online Boutique, Sock Shop, and Train Ticket datasets. 
%Due to space limit, the results on the Sock Shop dataset are available in our supplementary materials.
The fine-grained ground truths are derived from the indicators linked to the fault injection operations. For instance, when a CPU hog fault is injected into the order service, the fine-grained ground truth indicator is the "order-cpu" metric. 
We measure and report the accuracy for each fault type. Additionally, we perform statistical analysis using t-tests to check for pairwise differences among all RCA methods.

\textbf{(1) TORAI outperforms all baselines in fine-grained RCA}. In the Online Boutique dataset (Table \ref{tab:rq1-ob-fine}), TORAI achieves an average accuracy of 0.68 when diagnosing the root cause indicators. Meanwhile, BARO, the second-best method, achieves the average accuracy of 0.39. This is because BARO only uses hypothesis testing to infer the root cause while our TORAI also uses causal information and cluster the anomaly severity behaviour.

\textbf{(2) Hypothesis testing-based and causal inference-based methods deliver competitive fine-grained RCA performance.} BARO, CausalRCA, and CIRCA demonstrate strong fine-grained RCA capabilities. On the Train Ticket dataset, BARO achieves an average A5 score of 0.54 (the second highest), while our TORAI achieves a score of 0.68. Notably, BARO relies solely on hypothesis testing without considering causal relationships between time series and their anomaly severity. CausalRCA consistently performs well in fine-grained RCA, ranking among the top methods (TORAI, CIRCA, BARO). It constructs a causal graph using DAG-GNN and applies PageRank to identify root causes. CIRCA achieves an average A5 score of 0.67 on the Sock Shop dataset, while TORAI achieves a slightly higher score of 0.69. CIRCA combines causal inference with regression-based hypothesis testing. Notably, TORAI integrates causal inference in its CausalRanker, hypothesis testing in FineGrainer, and clustering in SymptomCluster, enabling it to achieve the best overall performance.

\subsection{RQ3. Efficiency in Root Cause Analysis} \label{sec:experimental-rq2}

\setlength{\columnsep}{10pt}
\begin{wraptable}{r}{6.5cm}
\vspace{-30pt}
\caption{Efficiency comparison.
}
\vspace{-5pt}
\label{tab:r2-efficiency-1}
\resizebox{\linewidth}{!}{ \setlength\tabcolsep{4pt}
\begin{tabular}{l|r|r|r}
\hline
\multicolumn{1}{c|}{Method} & \multicolumn{1}{c|}{\textbf{\begin{tabular}[c]{@{}c@{}}Online\\ Boutique\end{tabular}}} & \multicolumn{1}{c|}{\textbf{\begin{tabular}[c]{@{}c@{}}Sock\\ Shop\end{tabular}}} & \multicolumn{1}{c}{\textbf{\begin{tabular}[c]{@{}c@{}}Train\\ Ticket\end{tabular}}} \\ \hline \hline 
BARO & 0.01 & 0.01 & 0.01 \\
TraceRCA & 2.55 & - & 14.75 \\
RCD & 3.91 & 2.96 & 20.87 \\
PDiagnose & 3.24 & 1.26 & 61.53 \\
MicroRank & 77.56 & - & 91.04 \\
HeMiRCA & 62.08 & 55.38 & 206.02 \\
CIRCA & 4.65 & 3.5 & 312.77 \\
MicroCause & 177.09 & 129.13 & 3935.85 \\
CausalRCA & 179.45 & 397.12 & (*) \\ 
\hline \hline
\textbf{TORAI} & 12.9 & 15.63 & 20.59 \\ \hline 
\end{tabular} }

\vspace{5pt}
{\footnotesize (*) Exceeding the limit of 2 hours per case. \\ (-) Sock Shop's trace data is unavailable.
}\
\vspace{-10pt}
\end{wraptable}
\quad \textbf{(1) TORAI analyzes the root cause of failures in seconds.} TORAI takes an average of 12.9, 15.63, and 20.59 seconds to perform RCA on the Online Boutique, Sock Shop, and Train Ticket datasets, respectively. In comparison, MicroCause takes 177.09, 129.13, and 3935.85 seconds while RCD, %which is integrated into our CausalRanker, 
completes RCA in 3.91, 2.96, and 20.87 seconds, respectively.

\textbf{(2) TORAI's efficiency remains stable regardless of the system's scale.} TORAI takes an average of 12.9 seconds to perform RCA on the Online Boutique system (11 services), and 20.59 seconds on the Train Ticket system (64 services). In contrast, most baselines tend to slow down considerably when applied to larger systems. For instance, CausalRCA takes 180 seconds to handle Online Boutique but 20200 seconds to handle a case Train Ticket (100 times slower). This is because CausalRCA use deep neural network to perform causal discovery on all input time series, leading to a huge number of possible edges to analyse. BARO is the fastest method since it uses simple statistical method applying on time series data individually.

In summary, our TORAI can efficiently analyze root causes using multi-source telemetry data with minimal overhead, providing results in seconds even for microservice systems with a large number of services.

% \begin{wraptable}{r}{9cm}
\begin{table}[h]
\vspace{-4pt}
\caption{Ablation Study of TORAI}
\label{tab:r3-ablation}
\vspace{-4pt}
\resizebox{0.6\linewidth}{!}{ \setlength\tabcolsep{3pt}
\begin{tabular}{c|l|rrr|ccc|rrr}
\hline
\multirow{2}{*}{\begin{tabular}[c]{@{}c@{}}Data\\ Source\end{tabular}} & \multicolumn{1}{c|}{\multirow{2}{*}{Method}} & \multicolumn{3}{c|}{\textbf{Online Boutiq}} & \multicolumn{3}{c|}{\textbf{Sock Shop}} & \multicolumn{3}{c}{\textbf{Train Ticket}} \\ \cline{3-11} 
 &  & \multicolumn{1}{c}{\textit{T1}} & \multicolumn{1}{c}{\textit{T3}} & \multicolumn{1}{c|}{\textit{A5}} & \textit{T1} & \textit{T3} & \textit{A5} & \multicolumn{1}{c}{\textit{T1}} & \multicolumn{1}{c}{\textit{T3}} & \multicolumn{1}{c}{\textit{A5}} \\ \hline \hline
\multirow{3}{*}{\begin{tabular}[c]{@{}c@{}}Metric,\\Log\end{tabular}} & CausalRanker & 0.6 & 0.81 & 0.77 & \multicolumn{1}{r}{0.39} & \multicolumn{1}{r}{0.73} & \multicolumn{1}{r|}{0.66} & 0.09 & 0.2 & 0.17 \\
 & SeverityScorer & 0.8 & 0.93 & 0.91 & \multicolumn{1}{r}{0.8} & \multicolumn{1}{r}{0.95} & \multicolumn{1}{r|}{0.93} & 0.72 & 0.83 & 0.81 \\
 & FineGrainer & 0.69 & 0.82 & 0.81 & 0.07 & 0.96 & 0.76 & 0.69 & 0.82 & 0.81 \\
 & \textbf{TORAI} & 0.8 & 0.95 & 0.91 & \multicolumn{1}{r}{0.84} & \multicolumn{1}{r}{0.96} & \multicolumn{1}{r|}{0.94} & 0.74 & 0.84 & 0.83 \\ \hline
\multirow{3}{*}{\begin{tabular}[c]{@{}c@{}}Metric,\\Log,\\Trace\end{tabular}} & CausalRanker & 0.63 & 0.8 & 0.79 & - & - & - & 0.1 & 0.64 & 0.54 \\
 & SeverityScorer & 0.82 & 0.94 & 0.92 & - & - & - & 0.71 & 0.87 & 0.85 \\
 & FineGrainer & 0.69 & 0.82 & 0.81 & - & - & - & 0.69 & 0.82 & 0.81 \\
 & \textbf{TORAI} & 0.83 & 0.95 & 0.93 & - & - & - & 0.77 & 0.92 & 0.89 \\ \hline
\end{tabular} }

{\footnotesize (-) Sock Shop's trace data is unavailable.}
\vspace{-12pt}
\end{table}

%\subsection{RQ4. Effectiveness of TORAI’s Components} \label{sec:experimental-rq3}
\subsection{RQ4. Ablation Study} \label{sec:experimental-rq3}

In this section, we conduct an ablation analysis to assess the contribution of each constituent component to TORAI's overall performance. Table \ref{tab:r3-ablation} presents the effectiveness comparison of TORAI and its three components: SeverityScorer, CausalRanker, and FineGrainer, across three datasets. Specifically, SeverityScorer ranks the root cause services based on the average value of the severity vector [$\rho_m$,~$\rho_l$,~$\rho_{tc}$]. CausalRanker uses RCD \cite{Azam2022rcd} to rank the root cause services from multi-source time  
series data,~(Sec. \ref{sec:baselines}). Meanwhile, FineGrainer performs hypothesis testing on multi-source time series data and ranks root causes based on the magnitude of $\gamma$.

The ablation results from Table \ref{tab:r3-ablation} demonstrate that all constituent components have positive effects on TORAI's overall performance. For instance, on the Train Ticket dataset, TORAI achieves T1, T3, and A5 accuracies of 0.77, 0.92, and 0.89, respectively, while SeverityScorer achieves 0.71, 0.87, and 0.85, and FineGrainer achieves 0.69, 0.82, and 0.81. SeverityScorer relies on the anomaly severity score to rank the root causes without considering the causal structure between services, hence its performance is lower than TORAI. Similarly, FineGrainer also relies purely on the hypothesis testing results of the time series individually without taking into account the causal relationship among them, resulting in poorer performance compared to our TORAI. Meanwhile, in the same scenario, CausalRanker achieves scores of 0.1, 0.64, and 0.54, respectively. It has relatively poor performance by itself because it solely focuses on causal learning without considering the anomaly severity score.

\subsection{RQ5. How Does TORAI Perform in Real-World Scenarios?}

In this section, we further evaluate TORAI on 10 real-world failures collected from a production system. We also demonstrate the ability of TORAI in diagnosing code-level failures in Section~\ref{sec:code-level}.

\subsubsection{System and Data Description.}
The real-world failures are collected from a production microservice system of a major Internet service provider. The system consists of multiple components (e.g., load balancers, web/app servers, databases) categorized into five classes: OSB (Oracle Service Bus), service, DB, Docker, and OS. The system serves more than 50 million users. To monitor the system, engineers collect multi-source telemetry data. Due to confidentiality concerns, only metrics and traces are available, whereas logs are not collected. The dataset contains 10 failures with ground-truth labels for root cause components (e.g., services, databases). Each failure originates from a single component and propagates to others due to complex dependencies, causing their telemetry data to become abnormal.

\begin{wraptable}{r}{0.45\textwidth}
\vspace{-10pt}
\centering
\caption{Example of collected data.} \label{tab:metrics}
\vspace{-5pt}
{\footnotesize (a) Metrics.} \\
\vspace{2pt}
\resizebox{0.8\linewidth}{!}{%
\begin{tabular}{c|c|c}
\hline
\textbf{time} & \textbf{docker1\_cpu} & \textbf{docker1\_mem} \\ \hline \hline
17336 & 0.216 & 0.352 \\ 
17337 & 0.115 & 0.401 \\ 
17338 & 0.116 & 0.386 \\  
17339 & 0.118 & 0.398 \\  \hline
\end{tabular}%
}

\vspace{5pt}
{\footnotesize (b) Traces.} \\
\vspace{2pt}
\resizebox{\linewidth}{!}{%
\setlength\tabcolsep{1pt}
\begin{tabular}{l|l|l|l|c}
\hline
\textbf{time} & \textbf{id} & \textbf{service} & \textbf{callType} & \textbf{elapsedTime} \\ \hline \hline
17336 & cf8b.. & osb\_001 & OSB & 497 \\ 
17337 & 60cf.. & os\_021 & CSF & 102 \\ 
17338 & 4a93.. & docker\_003 & Remote.. & 1310 \\
17338 & fe23.. & product.. & ListProd.. & 56 \\ \hline
\end{tabular}%
}
\vspace{-5pt}
\end{wraptable}
Regarding the collected telemetry data, metrics are recorded at multiple layers. For service instances (i.e., Docker containers), 10 metrics are collected (e.g., thread\_total, fgct, session\_used, cpu\_used, mem\_used, etc.). For physical Linux host machines, 50 metric types are collected (e.g., Agent\_ping, Buffers\_used, Zombie\_Process, ss\_total, etc.). For Oracle databases, 47 database-specific metrics are collected (e.g., UndoTbs\_Pct, Used\_Tbs\_Size, User\_Commit, tnsping\_result\_time). The dataset contains over 169,000 traces in total. The data was collected since May 31, 2020, spanning 15 days. Each fault lasts for 5 minutes, and the root cause is confirmed by engineers. There are five fault types: CPU exhaustion, network delay, packet loss, container network error, and database failures. Fault locations are in databases, hosts, or service instances (containers). Table~\ref{tab:metrics} presents examples of metrics and trace data. Each column in the metric data represents a time series, and each row contains a value for a specific timestamp. Each trace has a unique ID, timestamp, corresponding service, call type, and elapsed time.

\subsubsection{Experimental setup}
In the collected dataset, logs are not available; we only have metrics and traces. We transform metrics and traces into time series as described in Sec.~\ref{sec:method-collect}. To ensure fairness, we feed the processed time series into TORAI along with four state-of-the-art RCA methods, namely BARO, RCD, CIRCA, and PDiagnose. We use AC@1, AC@3, AC@5, and Avg@5 as evaluation metrics to assess the ability to identify the root cause service of the failures.

% \begin{wraptable}{r}{0.5\textwidth}
\begin{table}
\centering
% \vspace{-10pt}
\caption{Real-world RCA Performance.}
\vspace{-7pt}
\label{tab:real-world-performance}
\resizebox{0.5\textwidth}{!}{%
\begin{tabular}{l|l|l|l|l}
\hline
Method & \textit{AC@1} & \textit{AC@3} & \textit{AC@5} & \textit{Avg@5} \\ \hline \hline
BARO~\cite{pham2024baro} & \textbf{0.70} & \underline{0.90} & \underline{0.90} & \underline{0.82} \\
CIRCA~\cite{Li2022Circa} & 0.20 & 0.50 & 0.70 & 0.46 \\
PDiagnose\cite{hou2021pdiagnose} & \underline{0.60} & 0.80 & 0.80 & 0.74 \\
RCD~\cite{Azam2022rcd} & 0.10 & 0.20 & 0.20 & 0.16 \\
\hline
\textbf{TORAI} & \underline{0.60} & \textbf{1.00} & \textbf{1.00} & \textbf{0.88} \\ \hline
\end{tabular}%
}
\vspace{-2pt}
\end{table}

\subsubsection{Results}
Table~\ref{tab:real-world-performance} presents the experimental results, demonstrating that TORAI outperforms state-of-the-art baselines. Notably, TORAI achieves 100\% in AC@3, meaning it can correctly rank the root cause within the top three recommendations with perfect accuracy. On average, TORAI attains an Avg@5 score of 0.88, outperforming BARO (0.82), PDiagnose (0.74), RCD (0.16), and CIRCA (0.46). RCD performs poorly because it processes all time series data indiscriminately, a limitation previously highlighted in~\cite{pham2024root}. Similarly, CIRCA constructs a causal graph using the PC algorithm on all time series data, leading to suboptimal performance. PDiagnose, which relies on traces to identify the root causes without incorporating metrics, achieves relatively strong results. Meanwhile, BARO employs hypothesis testing and delivers competitive performance; however, it does not account for causal relationships or severity symptoms. TORAI achieves the highest overall performance by effectively grouping abnormal services and conducting precise causal analysis, resulting in superior accuracy compared to the baselines. A replicable notebook is available in our replication package.

\begin{figure}[h]
\centering
\includegraphics[width=0.8\linewidth]{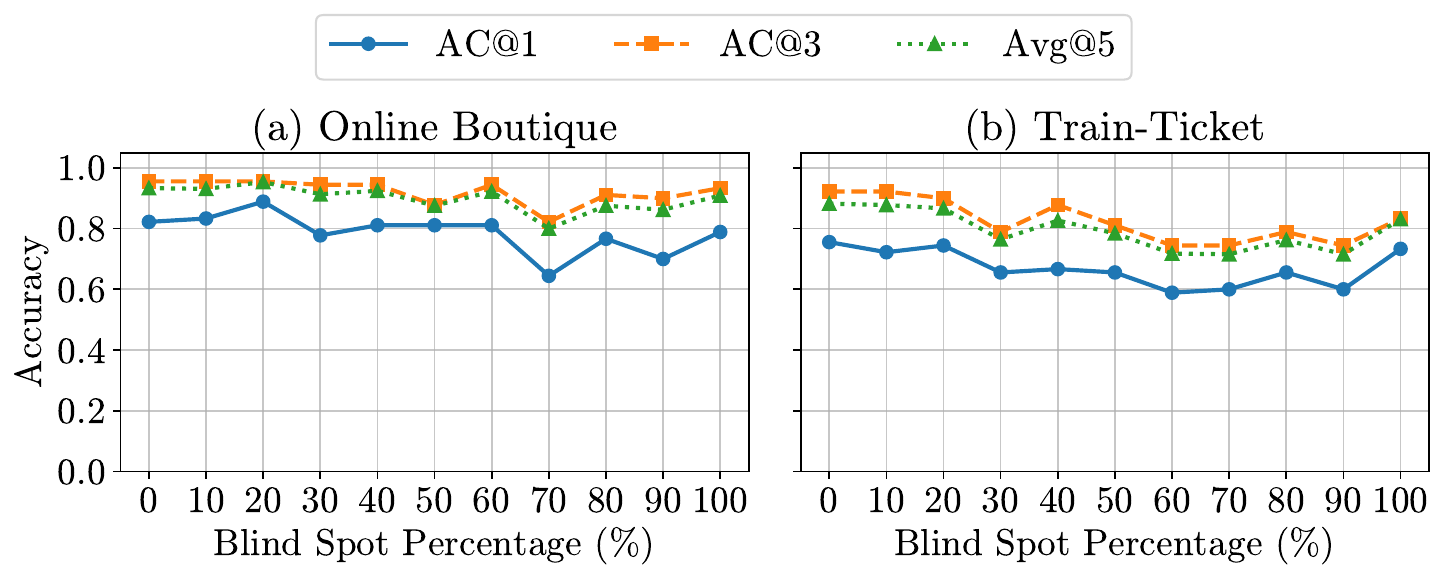}
\caption{Sensitivity analysis of TORAI performance under varying \textit{blind spot} percentages (i.e., proportion of services without trace data) on (a) Online Boutique and (b) Train Ticket systems. TORAI mostly maintains robust diagnostic accuracy (AC@1, AC@3, and Avg@5) across different blind spot levels.
} \label{fig:blindspot_sensitivity}
\vspace{-10pt}
\end{figure}

  % Online Boutique (OB) Results

  % | Blind Spot % | Top-1 | Top-3 | Avg@5 |
  % |--------------|-------|-------|-------|
  % | 0%           | 83.0% | 95.0% | 93.0% |
  % | 10%          | 83.3% | 95.6% | 93.1% |
  % | 20%          | 88.9% | 95.6% | 95.3% |
  % | 30%          | 77.8% | 94.4% | 91.3% |
  % | 40%          | 81.1% | 94.4% | 92.4% |
  % | 50%          | 81.1% | 87.8% | 87.6% |
  % | 60%          | 81.1% | 94.4% | 92.2% |
  % | 70%          | 64.4% | 82.2% | 80.0% |
  % | 80%          | 76.7% | 91.1% | 87.6% |
  % | 90%          | 70.0% | 90.0% | 86.2% |
  % | 100%         | 80.0% | 95.0% | 91.0% |

  % Train-Ticket (TT) Results

  % | Blind Spot % | Top-1 | Top-3 | Avg@5 |
  % |--------------|-------|-------|-------|
  % | 0%           | 77.0% | 92.0% | 89.0% |
  % | 10%          | 72.2% | 92.2% | 87.8% |
  % | 20%          | 74.4% | 90.0% | 86.7% |
  % | 30%          | 65.6% | 78.9% | 76.4% |
  % | 40%          | 66.7% | 87.8% | 82.7% |
  % | 50%          | 65.6% | 81.1% | 78.4% |
  % | 60%          | 58.9% | 74.4% | 71.8% |
  % | 70%          | 60.0% | 74.4% | 71.6% |
  % | 80%          | 65.6% | 78.9% | 76.2% |
  % | 90%          | 60.0% | 74.4% | 71.6% |
  % | 100%         | 74.0% | 84.0% | 83.0% |

\subsection{How Robust Is TORAI Under Varying Blind Spots?} \label{sec:sensitivity-blindspots}

To evaluate TORAI's sensitivity to varying levels of blind spots (i.e., trace unavailability), we conduct a sensitivity analysis by randomly removing traces from 0\% to 100\% of the instrumented services in 10\% increments on the Online Boutique and Train Ticket systems\footnote{The Sock Shop system has no traces by default (i.e., 100\% blind spots).}. Figure~\ref{fig:blindspot_sensitivity} presents TORAI's performance across these blind spot levels. We draw the following observations:

\textbf{(1) TORAI remains robust on the Online Boutique system.} As shown in Figure~\ref{fig:blindspot_sensitivity}(a), AC@1 ranges from 64.4\% to 88.9\% across all blind spot levels, with AC@3 remaining consistently above 82\%. Notably, even at 100\% blind spots (i.e., no trace data available), TORAI achieves 80\% AC@1 and 95\% AC@3, demonstrating that metrics and logs alone provide sufficient diagnostic signals for this system.

\textbf{(2) TORAI shows graceful degradation on the larger Train Ticket system.} Figure~\ref{fig:blindspot_sensitivity}(b) reveals that the 64-service Train Ticket system exhibits more sensitivity to blind spots, with Avg@5 dropping from 89\% (at 0\%) to a minimum of 71.6\% (at 70\%). However, the degradation is gradual rather than catastrophic. Interestingly, performance recovers at 100\% blind spots (74\% AC@1, 84\% AC@3). %, which we attribute to TORAI's multi-source design. 
We found that, when traces are entirely unavailable, the method relies purely on metrics and logs without the potential noise introduced by incomplete or partial trace information.

\textbf{(3) AC@3 is more stable than AC@1 across both systems.} While AC@1 fluctuates more noticeably, AC@3 remains relatively high (above 74\% for Train Ticket and above 82\% for Online Boutique) across all blind spot levels. This indicates that the true root cause is consistently ranked within the top three candidates, which is practical for engineers who can efficiently investigate a small set of services.

These findings are further supported by TORAI's strong performance on the Sock Shop system (Table~\ref{tab:rq1-ss}), which has 100\% blind spots by default and where TORAI achieves 0.84 AC@1 and 0.94 Avg@5. In summary, TORAI's multi-source design enables reliable RCA even under the presence of blind spots, addressing a key practical concern for systems where full distributed tracing instrumentation is infeasible.

\subsection{Diagnosing code-level faults} \label{sec:code-level}

\begin{wrapfigure}{r}{0.50\linewidth}
\vspace{-25pt}
\begin{lstlisting}[basicstyle=\ttfamily\scriptsize, backgroundcolor=\color{gray!10}, frame=single]
info: Grpc.AspNetCore.Server.CallHandler[7]
      Error status code 'FailedPrecondition'
      with detail 'Can't access cart storage.
      System.OverflowException: Value was either
      too large or too small for an Int32.
    at System.Number.ThrowOverflowException()
    at cartservice.cartstore.RedisCartStore
       .AddItemAsync(String, String, String)
       in /RedisCartStore.cs:line 54' raised. 
\end{lstlisting}
\caption{Stack traces as Fine-grained root cause of the code-level faults indicating line 54 of file RedisCartStore.cs is the root cause.}\label{fig:stack-traces}
\vspace{-10pt}
\end{wrapfigure}
While we follow established practice to benchmark on resource and network-related faults, TORAI can diagnose other fault types (e.g., code-level faults~\cite{cotroneo2019bad}) as long as the issues manifest symptoms in telemetry data. For example, if a code-level fault produces observable symptoms, such as exception logs or erroneous traces, our method can detect the root cause services with these indicators to assist engineers in diagnosing the actual fault more efficiently.

\begin{wrapfigure}{r}{0.58\textwidth}
\vspace{-5pt}
\centering
\includegraphics[width=0.9\linewidth]{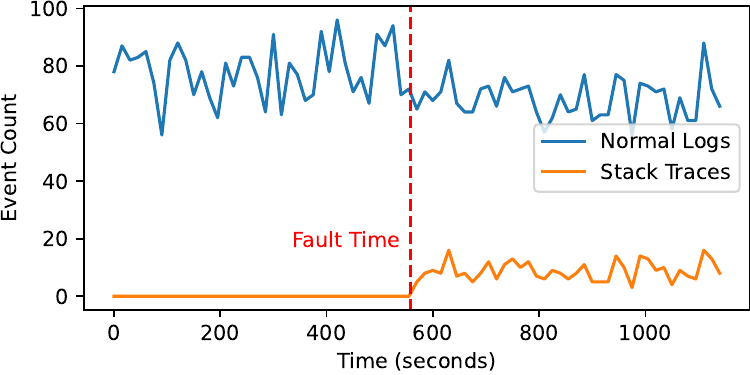}
\vspace{-10pt}
\caption{The frequency of normal logs/stack traces of cartservice.} \label{fig:cart-stack-trace}
\vspace{-10pt}
\end{wrapfigure}
To demonstrate this capability, we modified the \textit{cartservice} in the Online Boutique system to inject a code-level fault, namely \textit{Incorrect parameter values}, as described in~\cite{cotroneo2019bad}. Empirical studies show that incorrect parameter values are among the five most common faults in real-world projects and that injected faults can realistically simulate actual software faults~\cite{cotroneo2019bad}. After injecting this fault, the cartservice became unstable, resulting in failures in all its callers. The front-end services reported error codes, and the resource usage of the cartservice spiked. Other services, such as recommendation and shipping, were also affected. The root cause was traced back to the cartservice (RedisCartStore.cs:54), with the fine-grained root cause indicator being the stack trace presented in Figure~\ref{fig:stack-traces}.

Our FineGrainer precisely identifies the stack trace in the cartservice as a fine-grained root cause indicator, as its frequency during the failure period deviates significantly from the expected median, which is 0. The frequency of normal logs and stack traces for the cartservice is shown in Figure~\ref{fig:cart-stack-trace}. This demonstration is included in our replication package.

\section{Threats to Validity}

We now discuss threats to the validity of our study and the means we undertook to mitigate these threats. The \textit{internal threat} concerns the implementation, where bugs may affect the reliability of the results. To address this, we reused the public code for the baselines and performed experiments to replicate their results, ensuring their correctness. To avoid the influence of randomness, we repeated the experiments five times and reported the average results together with the statistical analysis. The \textit{construct threat} concerns the evaluation metrics. To address this, we used standard evaluation metrics extensively employed in the literature to evaluate the performance of RCA methods~\cite{Azam2022rcd, pham2024baro, pham2024root}. The \textit{external threat} concerns the deployment of microservice applications and data collection strategies. To address this, we follow established practice to deploy these systems and collect data, as described in Section \ref{sec:dataset}. These systems are widely recognized in academia for testing microservices-related methods~\cite{Jinjin2018Microscope, Azam2022rcd, wu2021microdiag, Xin2023CausalRCA, wu2022automatic, he2022graph, yu2021microrank, zhou2018trainticket, Wang2021evalcausal}. Another potential threat concerns the assumptions underlying our method. We assume that root cause indicators exhibit significant distributional changes, which is supported by prior works~\cite{pham2024baro, Shan2019Ediagnosis} and recent theoretical analysis~\cite{neurips2025rca}. We do not assume the root cause service always exhibits the highest raw anomaly score, instead, severity-based clustering serves as a coarse candidate selection mechanism, with final decisions made by CausalRanker and FineGrainer.

The \textit{conclusion threat} is tied to the fault types as microservices can experience various faults \cite{mariani2018localizing}. We acknowledge that different software applications and faults could have different properties and failure propagation mechanisms, which could impact the conclusions in this paper. However, we believe that these fault types are representative since they have been used in many previous studies~\cite{Jinjin2018Microscope, Azam2022rcd, wu2021microdiag, Xin2023CausalRCA, he2022graph, dan2021tracerca, zhou2018trainticket}. Furthermore, our method is specifically designed for microservices systems; however, if a software system has multiple components interacting with each other, our method can be adapted to identify the root cause of failures in these systems. Expanding TORAI to work with other types of systems, such as distributed database systems, could be a potential future work. There may be other threats related to the underlying tools, our extracted data, that we have not considered here. To enable exploration of these potential threats and to facilitate replication and extension of our work, we make available our tools and data.

\section{Conclusion}

In this work, we propose TORAI, a novel unsupervised fine-grained RCA method that leverages multi-source telemetry data to accurately identify both coarse-grained root cause services and fine-grained root cause indicators of failures in microservice systems. The novelty of our proposed TORAI lies in the effective combination of unsupervised techniques, including clustering, causal analysis, and hypothesis testing, which enables the integration of multi-source telemetry data for RCA without requiring full trace coverage or labeled data, addressing limitations of existing RCA approaches. Extensive experiments on three benchmark systems and real-world failures demonstrate TORAI's superiority in both effectiveness and efficiency compared to existing methods.

\section{Data Availability}
We have integrated TORAI into our open-source benchmark RCAEval~\cite{pham2025rcaeval}, which can be accessed on GitHub at \url{https://github.com/phamquiluan/rcaeval}. Additionally, an immutable artifact for TORAI is available on Figshare~\cite{torai_sourcecode}, together with the experimental datasets~\cite{torai_datasets}.

\begin{acks}
This research was supported by the Australian Research Council Discovery Project (DP220103044), Google Cloud Credit for Research, and RMIT Race Hub via the RMAS scheme. 
\end{acks}

\bibliographystyle{ACM-Reference-Format}
\bibliography{reference}
\end{document}